\documentclass[12pt]{article}
\usepackage{amsmath,amssymb}
\usepackage{graphicx,url}

\makeatletter
\newif\if@preliminary
\@preliminaryfalse
\def\preliminary{\@preliminaryfalse}

\def\bq{\begin{equation}}
\def\eq{\end{equation}}
\def\ba{\begin{eqnarray}}
\def\ea{\end{eqnarray}}

\def\preprintno#1{\def\@preprintno{#1}}
\def\address#1{\def\@address{#1}}
\def\email#1#2{\thanks{\tt #1@{}#2}}
\def\abstract#1{\def\@abstract{#1}}
\renewcommand\abstractname{ABSTRACT}
\newlength\preprintnoskip
\newlength\abstractwidth
\@titlepagetrue
\renewcommand\maketitle{\begin{titlepage}
  \let\footnotesize\small
  \hfill\parbox{\preprintnoskip}{
  \begin{flushright}\@preprintno\end{flushright}}\hspace*{1cm}
  \vskip 60\p@
  \begin{center}
    {\Large\bf\boldmath \@title \par}\vskip 1cm
    {\sc\@author \par}\vskip 3mm
    {\@address \par}
    \if@preliminary
      \vskip 2cm {\large\sf PRELIMINARY DRAFT \par \@date}
    \fi
  \end{center}\par
  \@thanks
  \vfill
  \begin{center}
    \parbox{\abstractwidth}{\centerline{\abstractname}
    \vskip 3mm
    \@abstract}
  \end{center}
  \end{titlepage}
  \setcounter{footnote}{0}
  \let\thanks\relax\let\maketitle\relax
  \gdef\@thanks{}\gdef\@author{}\gdef\@address{}
  \gdef\@title{}\gdef\@abstract{}\gdef\@preprintno{}
}

\def\@citex[#1]#2{\if@filesw\immediate\write\@auxout{\string\citation{#2}}\fi
  \def\@citea{}\@cite{\@for\@citeb:=#2\do
    {\@citea\def\@citea{,\penalty\@m}\@ifundefined
       {b@\@citeb}{{\bf ?}\@warning
       {Citation `\@citeb' on page \thepage \space undefined}}%
\hbox{\csname b@\@citeb\endcsname}}}{#1}}
\def\citerange{\@ifnextchar [{\@tempswatrue\@citexr}{\@tempswafalse\@citexr[]}}
\def\@citexr[#1]#2{\if@filesw\immediate\write\@auxout{\string\citation{#2}}\fi
  \def\@citea{}\@cite{\@for\@citeb:=#2\do
    {\@citea\def\@citea{--\penalty\@m}\@ifundefined
       {b@\@citeb}{{\bf ?}\@warning
       {Citation `\@citeb' on page \thepage \space undefined}}%
\hbox{\csname b@\@citeb\endcsname}}}{#1}}
\long\def\@makecaption#1#2{
  \vskip\abovecaptionskip
  \sbox\@tempboxa{#1: \emph{#2}}
  \ifdim \wd\@tempboxa >\hsize
    #1: \emph{#2}\par
  \else
    \hbox to\hsize{\hfil\box\@tempboxa\hfil}
  \fi
  \vskip\belowcaptionskip}

\def\fmslash{\@ifnextchar[{\fmsl@sh}{\fmsl@sh[0mu]}}
\def\fmsl@sh[#1]#2{
  \mathchoice
    {\@fmsl@sh\displaystyle{#1}{#2}}
    {\@fmsl@sh\textstyle{#1}{#2}}
    {\@fmsl@sh\scriptstyle{#1}{#2}}
    {\@fmsl@sh\scriptscriptstyle{#1}{#2}}}
\def\@fmsl@sh#1#2#3{\m@th\ooalign{$\hfil#1\mkern#2/\hfil$\crcr$#1#3$}}
\makeatother

\newcommand{\pd}{\partial}
\newcommand{\LL}{\mathcal{L}}

\newcommand{\Tr}{\mathop{\rm Tr}}
\newcommand{\diag}{\mathop{\rm diag}}
\newcommand{\ii}{{\rm i}}
\newcommand{\Op}{\mathcal{O}}

\newcommand{\hc}{\text{h.c.}}

\renewcommand{\Re}{\text{Re}}
\renewcommand{\Im}{\text{Im}}
\newcommand{\pl}{\mathcal{P}_L}
\newcommand{\pr}{\mathcal{P}_R}

\newcommand{\MeV}{{\ensuremath\rm MeV}}
\newcommand{\GeV}{{\ensuremath\rm GeV}}
\newcommand{\TeV}{{\ensuremath\rm TeV}}

\newcommand{\fb}{{\ensuremath\rm fb}}
\newcommand{\ab}{{\ensuremath\rm ab}}

\begin{document}

\preliminary        


\preprintno{DESY 04-137, TTP 04-16, NSF-KITP-04-124\\hep-ph/0411213\\[0.5\baselineskip] November 16, 2004}

\title{PSEUDO-AXIONS\\ IN LITTLE HIGGS MODELS}

\author{
 W.~Kilian\email{wolfgang.kilian}{desy.de}$^a$,
 D.~Rainwater\email{rain}{pas.rochester.edu}$^b$,
 and J.~Reuter\email{juergen.reuter}{desy.de}$^{a,c}$
}

\address{\it
$^a$Deutsches Elektronen-Synchrotron DESY, D--22603 Hamburg, Germany
\\[.5\baselineskip]
$^b$Department of Physics and Astronomy, University of Rochester,\\
Rochester, NY 14627, USA
\\[.5\baselineskip]
$^c$Institut f\"ur Theoretische Teilchenphysik, Universit\"at Karlsruhe,\\
D--76128 Karlsruhe, Germany
}

\abstract{
Little Higgs models have an enlarged global symmetry which makes the
Higgs boson a pseudo-Goldstone boson.  This symmetry typically
contains spontaneously broken $U(1)$ subgroups which provide light
electroweak-singlet pseudoscalars.  Unless such particles are absorbed
as the longitudinal component of $Z'$ states, they appear as
pseudoscalars in the physical spectrum at the electroweak scale.  We
outline their significant impact on Little Higgs phenomenology and
analyze a few possible signatures at the LHC and other future
colliders in detail.  In particular, their presence significantly
affects the physics of the new heavy quark states predicted in Little
Higgs models, and inclusive production at LHC may yield impressive
diphoton resonances.}

\maketitle


\section{Introduction}

Little Higgs models~\cite{LHorig} have recently emerged as an
alternative solution to the naturalness problem of the Higgs sector in
the Standard Model (SM).  In these models, the origin of the
electroweak (EW) scale is identified as new dynamics in the multi-TeV
range that involves the spontaneous breaking of some extra continuous
symmetry.  While the exact nature of this dynamics, the UV completion,
is undetermined --- scenarios involving strong
interactions~\cite{UV-strong}, iterated Little Higgs
models~\cite{UV-iterated}, or supersymmetry~\cite{UV-SUSY} have been
proposed --- the low-energy effective theory below the UV scale,
$\Lambda$, is determined by the assumed symmetries.  Various
realizations of the Little Higgs symmetry structure have been
proposed~\cite{LHmin,minmoose,simple,LHmod,simplest}.

If at the scale $\Lambda$ an exact global (i.e., ungauged) symmetry is
spontaneously broken, the spectrum contains Nambu-Goldstone bosons
(NGBs) which are massless and have no renormalizable interactions.  In
analogy to low-energy QCD, we denote the would-be decay constant of
these scalars by~$F$.  Naive dimensional analysis relates this scale
to the cutoff $\Lambda$ by $F\sim\Lambda/4\pi$.

In Little Higgs models, the SM Higgs doublet is among these NGBs.
This would be a natural explanation for a weakly interacting Higgs
sector.  However, since the Higgs doublet does have nontrivial
renormalizable interactions --- gauge, Yukawa, and self-couplings ---
there must be interactions in the initial Lagrangian which break the
global symmetry explicitly.  In general, such symmetry-breaking
interactions introduce a one-loop Coleman-Weinberg
potential~\cite{CW73} and a Higgs mass proportional to
$m_H^2\sim\Lambda^2/(16\pi^2)\sim F^2$.  Without fine-tuning the
parameters or adding additional fields, one then derives $F\sim v$,
i.e., the new symmetry-breaking scale is near the electroweak symmetry
breaking (EWSB) scale~\cite{H-Goldstone}.

The new ingredient in Little Higgs models is the mechanism of
collective symmetry breaking, as it was observed in the context of
deconstructed extra-dimension models~\cite{LH-deconst}.  Each
renormalizable scalar interaction breaks the postulated global
symmetry explicitly, but leaves a continuous subgroup intact.
Spontaneous breaking of this remaining global symmetry then still
implies the existence of a massless NGB.  However, all interactions
together break all global symmetries explicitly, and no particle stays
massless to all orders.  The Higgs doublet is found among the scalars
that still remain NGBs as long as only a single symmetry-breaking
coefficient (spurion) is present in the Lagrangian, but acquire masses
once all spurions are turned on.  For instance, for two spurions
$g_1,g_2$ the resulting quadratic term in the effective Higgs
potential is proportional to $(g_1/4\pi)^2(g_2/4\pi)^2\Lambda^2$.  As
usual, EWSB is triggered by such a term, and thus we have a
three-scale model with
\bq
v \sim F/4\pi \sim \Lambda/(4\pi)^2 \; .
\eq

Since the UV-completion scale $\Lambda$ is parameterically two orders
of magnitude above the EW scale, any sign of the associated dynamics
is strongly suppressed, and we are left with the virtual effects of
new particles at the intermediate scale $F$.  For a consistent
implementation of collective symmetry breaking, enlarged symmetries
must be introduced in all sectors of the model, so we expect new
vector, spinor, and scalar particles with masses of order $F$.  The
low-energy traces of these particles can be computed and have been
used to derive limits on the parameter space of any given Little Higgs
model~\cite{Csaki,Hewett,Han,Csaki2,PPP03,CDO03,KR03}.

In this paper we study the phenomenological consequences of a
particular property of the Little Higgs mechanism.  To allow for an EW
doublet among the NGBs, the spontaneous breaking of the global
symmetry group typically involves a reduction of the group rank,
e.g. $SU(3)\to SU(2)$.  While the Higgs doublet in this example
corresponds to the off-diagonal broken generators (analogous to the
kaons in QCD), there is also one broken diagonal generator (analogous
to the $\eta$).  If there were no explicit symmetry-breaking terms,
this particle, which acts as a pseudoscalar in its fermionic
couplings, would behave as an axion, so we may call it the
pseudo-axion of a Little Higgs model.  Similar particles show up in
other models which involve enlarged global symmetries, e.g.
technicolor models~\cite{Technicolor,Technicolor-axion}, see-saw
topcolor models~\cite{Topcolor,Topcolor-axion}, or the
NMSSM~\cite{NMSSM,NMSSM-axion}.  To determine the detailed model
structure, one must experimentally establish any pseudo-axions as part
of the Higgs sector and determine their properties.  We find that
these states have significant, broad impact on Little Higgs
phenomenology.


\section{Pseudo-axions in Little Higgs Models}

In the bosonic sector, two different lines of model-building realize
collective symmetry breaking.  In models built along the lines of {\it
theory-space} or {\it moose}
models~\cite{LHorig,minmoose,simple,LHmod,simplest}, the
global-symmetry representation is reducible, i.e., in the scalar
sector there are several distinct multiplets with gauge quantum
numbers.  The gauge coupling of any multiplet acts as a spurion which
reduces the global symmetry down to the exactly realized gauge
symmetry.  This is spontaneously broken down to the EW gauge group
$SU(2)\times U(1)$, and some NGBs are absorbed as the longitudinal
components of new heavy vector bosons.  However, if there are several
scalar multiplets, there are not enough gauge bosons to absorb all the
NGBs.  The masses of the uneaten linear combinations are proportional
to two or more spurions (i.e., gauge couplings) and do not appear up
to one-loop order.  Introducing scalar self-couplings as additional
spurions, the structure becomes more complicated and more scalar
multiplets are needed to keep the Little Higgs mechanism working, but
the line of reasoning remains unchanged.

For concrete examples, let us consider the {\it Minimal
Moose}~\cite{minmoose} and {\it Simple Group}~\cite{simple} models.
In the first, with exactly one scalar coupling turned on, the global
symmetry breaking is $[SU(3)]^4\to SU(3)$.  The group rank is reduced
by $k=6$ units, yielding $6$ pseudo-axion candidates.  The rank of the
spontaneously broken gauge group $SU(3)\times SU(2)\times U(1)$ is
$r=2$ units larger than the SM EW group, so $2$ axions are eaten to
make vectors massive and $k-r=4$ of them remain in the low-energy
spectrum. (Two are EW singlets, while the other two are the neutral
members of two real EW triplets.)  Similarly, in the Simple Group
model, if exactly one scalar coupling is turned on, the pattern of
global symmetry breaking is $[SU(4)]^3\to [SU(3)]^2\times SU(2)$,
while the gauge group is $SU(4)\times U(1)$.  This yields $k=4$ and
$r=2$, so there are $k-r=2$ pseudo-axions in this case.

As an alternative, some models implement an irreducible representation
of the symmetry group in the scalar sector~\cite{LHmin,LHmod}.  In
this case, to provide independent spurions in the gauge sector, the
enlarged gauge symmetry must be such that part of it commutes with the
EW gauge group.  In this setting, for which the {\it Littlest Higgs}
model~\cite{LHmin} serves as an example, scalar self-couplings arise
at one loop from integrating out the heavy vector bosons and fermions.
Here, the global symmetry breaking is $SU(5)\to SO(5)$, with rank
reduction $k=2$.  If the extra gauge group is $SU(2)$, one
pseudo-axion ($r=1$) is eaten in its symmetry breaking, and one
($k-r=1$) remains in the spectrum.  On the other hand, if the extra
gauge group is $SU(2)\times U(1)$, as originally proposed, both axions
disappear from the spectrum.

This example demonstrates that the new particles become unphysical if
all extra broken $U(1)$ symmetries are gauged.  However, in that case
we expect the corresponding number of new $Z'$ bosons with masses of
order~$F$.  These states are generally easy to detect at future
colliders, either directly as resonances in $q\bar{q}$ or $e^+e^-$
annihilation~\cite{Han,BPP03}, or indirectly via the observation of
contact interactions and of mixing with the standard $Z$
boson~\cite{Csaki,Hewett}.  In some models, these effects can be used
to rule out much of the parameter space from existing data
alone~\cite{Csaki,Hewett,Han,Csaki2,PPP03,CDO03,KR03}.  In the
following, we therefore consider the situation where the extra groups
are ungauged~\cite{Han,Csaki2,PPP03} and the pseudo-axions are
physical.


\section{Pseudo-axion Interactions}
\label{sec:axion}

Let us consider the case of a single pseudo-axion which we denote
by~$\eta$.  As an EW singlet it does not couple to EW gauge
bosons at leading order.  Furthermore, if $\eta$ is an exact NGB, it
has no potential and does not couple to Higgs bosons either.  However,
in the fermion sector the situation is different.  To account for
Higgs Yukawa couplings, Little Higgs models contain interactions of
fermions with the NGB multiplet, inducing Yukawa couplings for the
axion.  If the $U(1)$ symmetry generated by $\eta$ is parameterized by
\bq
\xi = \exp\frac{i}{F}\eta,
\eq
where $F$ is the symmetry breaking scale of the Little Higgs model, 
each field $\psi_i$ transforms like $\psi_i\to\xi^{\beta_i}\psi_i$,
where $\beta_i$ is the corresponding $U(1)$ charge.

In some models, for some of the axions, these couplings are fixed by
the symmetry structure that generates EWSB and can thus be deduced
from the study of observables in the gauge, Higgs, or fermion sectors.
In other models they cannot be determined this way.  Typically, the
axion interactions depend on additional parameters, so their
observation provides independent information on the high-energy
structure of the model.  In the following, we present three specific
models where axions play a role and discuss their interactions in some
detail.


\subsection{Mass and decays}

For a phenomenological discussion of an $\eta$ axion, we need an
estimate of its mass.  If $\eta$ is an exact NGB, it would be exactly
massless.  This appears to be the case in some of the models we
discuss in the following subsections.  However, even then the
$U(1)_\eta$ symmetry must be explicitly broken at some stage, so that
$\eta$ picks up a mass; a massless particle coupled to fermions (even
if this coupling is suppressed by $v/F$ and CKM factors) would induce
a long-range force many orders of magnitude stronger than gravity and
is therefore very strongly ruled out.

In general, the $U(1)_\eta$ symmetry will be anomalous.  Then, if
$m_\eta$ is below the QCD scale, $\eta$ becomes a Peccei-Quinn axion,
which is ruled out for the parameter range of interest.  Hence, we can
assume a lower mass limit in the $\GeV$ range.

In any model, we can also put a rough upper limit on $m_\eta$ by
considering its influence on the Higgs mass.  To this end, we first
note that a vertex $\eta^2 H^2$ in the scalar potential would
re-introduce a quadratic divergence in the Higgs mass at one loop,
i.e., a correction $\Delta m_H^2\sim\Lambda^2/(16\pi^2)\sim F^2$.  If
there is such a term, its coefficient should be parameterically of
order $1/(16\pi^2)$, so that the induced Higgs mass correction is at
most of order $v^2$.  Next consider $\eta H$ scattering: the NGB
Lagrangian will contain derivative interactions of the form
$\pd\eta^2\pd H^2/F^2$ which at one-loop order yield an effective
$\eta^2H^2$ vertex with a quadratic divergence proportional to
$(m_\eta^2/F^4)\Lambda^2/(16\pi^2)\sim m_\eta^2/F^2$.  To keep this
consistent without fine-tuning, we generally have to require
$m_\eta\lesssim v$, i.e., the EW scale is an upper limit for the
pseudo-axion mass.

All $\eta$ couplings to SM particles are suppressed by $v/F$, hence
the rates for direct production are generically a factor of $(v/F)^2$
smaller than the corresponding Higgs production channels.  For the
decays, throughout the mass range we expect a similar pattern as for a
{\it light} Higgs boson: dominant ($t\bar{t}$,) $b\bar{b}$,
$\tau^+\tau^-$, \ldots\ branching ratios (BRs), depending on kinematic
accessibility, and some fraction of gauge-boson pairs, i.e., $gg$,
$\gamma\gamma$, and $Z^{(*)}\gamma$, $ZZ^{(*)}$, $WW^{(*)}$.  Since
the axion is CP-odd, the vector-boson pair branching fractions are
loop-induced and therefore stay small even above the thresholds for
on-shell $WW$ and $ZZ$ pair production.


\subsection{The Littlest Higgs model}
\label{sec:littlest}

Here we derive the pseudo-axion interactions in the Littlest Higgs
model~\cite{LHmin} with only one $U(1)$
gauged~\cite{Han,Csaki2,PPP03}.  While this model has been extensively
discussed in the literature, the pseudo-axion interactions have so far
been ignored.

In this model, the NGB multiplet is parameterized by a $5\times 5$
matrix
\bq
\Xi = \left(\exp\frac{2i}{F}\Pi\right)\Xi_0 \; , 
\eq
where
\bq
\Xi_0 =
  \begin{pmatrix}
    0 & 0 & 1_{2\times 2} \\ 0 & 1 & 0 \\ 1_{2\times 2} & 0 & 0
  \end{pmatrix}
\quad {\rm and} \quad
\Pi = \frac{1}{\sqrt2}
  \begin{pmatrix}
    \eta/\sqrt{10} & h & \phi \\ 
    h^\dagger & -4\eta/\sqrt{10} & h^T \\ 
    \phi^\dagger & h^* & \eta/\sqrt{10}
  \end{pmatrix} \; .
\eq
We have included only the physical states: the Higgs doublet, the
$\eta$ singlet which multiplies the (canonically normalized) generator
$\diag(1,1,-4,1,1)/2\sqrt{10}\,$, and $\phi$ which is a complex
triplet written as a symmetric $2\times 2$ matrix.  The triplet
becomes heavy [i.e., its mass is of order $F$] and has little impact
on low-energy phenomenology, so we ignore it in the following.

The kinetic term for the NGB multiplet is given by
\bq\label{kinetic}
  \frac{F^2}{8} \Tr[ (D_\mu \Xi)^* (D^\mu \Xi)]
  \, = \, |D h|^2 + \frac12 (\pd\eta)^2 + \cdots \, ,
\eq
which fixes the $\eta$ field normalization.  In writing the Yukawa
interaction we have to allow for $\eta$-dependent factors
\bq
\xi = \exp\frac{i}{\sqrt{5}\,F}\eta \; ,
\eq
where the normalization has been adjusted for later convenience.  The
third-generation fermions are the left-handed quark doublet
$q_L=(t_L,b_L)^T$, the right-handed singlet $t_R$, and the new
singlets $T_R,T_L$.  We also define the matrices
\bq
\chi_L =
  \begin{pmatrix}
    i\tau^2 \xi^{\beta_0}T_L & i q_L & 0 \\
    -i q_L^T & 0 & 0 \\ 0 & 0 & 0
  \end{pmatrix}
\quad {\rm and} \quad
i T_2^2 \, = \, \diag(0,0,-i\tau^{2\ast})/2 \; .
\eq
With these definitions, the Yukawa interaction has the
form~\cite{LHmin,KR03}
\bq\label{yukawa}
\LL_Y^t = \lambda_1 F \xi^{\beta_1}
          \bar{t}_R\Tr\left[\Xi^* (iT_2^2) \Xi^* \chi_L\right]
          - \lambda_2 F \xi^{\beta_2} \bar{T}_R T_L
          + \hc
\eq
The parameters $\beta_{0,1,2}$ are real numbers determined by the
differences of fermion $U(1)_\eta$ charges.  The model has no
prediction for these numbers (note that anomaly cancellation is not an
issue since the $U(1)_\eta$ symmetry may well be anomalous), so we
leave them as free parameters.  In particular, we allow for
$\beta_2\neq 0$.  If this is the case, the $T$ quark mass is protected
by a chiral symmetry and is therefore naturally of order $F$, the
$U(1)_\eta$ breaking scale.\footnote{Previous papers~\cite{Han,Csaki2}
assumed that the $T$ quark is vectorlike, leaving this fine-tuning
problem unsolved.}

When the Yukawa interaction in Eq.~(\ref{yukawa}) is expanded in
powers of $1/F$, it yields a mass term for the new fermion $T$, a
mixing of $t$ and $T$, a Higgs Yukawa interaction of the top quark,
and higher-order terms:
\begin{align}\label{yuk-exp1}
\begin{split}
\LL_Y^t & =
  - \lambda_1 F\bar{t}_R T_L - \lambda_2 F\bar{T}_R T_L
  + \sqrt2\,\lambda_1 \bar{t}_R h^T\tilde q_L \\
& \quad
  + \frac{i}{\sqrt5}(2-\beta_0-\beta_1) \lambda_1 \bar{t}_R\eta T_L
  - \frac{i}{\sqrt5} \beta_2 \lambda_2 \bar{T}_R\eta T_L
  + \cdots + \hc \; ,
\end{split}
\end{align}
where $\tilde{q}_L = i\tau^2 q_L$. 

The mass matrix is diagonalized first by a rotation of the
right-handed fields with the parameters
\bq
s = \lambda_1/\lambda \, , \qquad
c = \lambda_2/\lambda \, , \qquad
\lambda = \sqrt{\lambda_1^2 + \lambda_2^2} \; ,
\eq
which are expected to be ${\cal O}(1)$, and then by subsequent
rotations of the left- and right-handed fields where the rotation
angles are suppressed by powers of $v/F$.  Keeping only the leading
terms, we write the interactions in terms of the physical fields $H$
and $\eta$ and the quark masses $M_T=\lambda F$ and $m_t=sc\lambda v$,
where $v$ is the Higgs vacuum expectation value:
\begin{align}\label{yuk-exp2}
\begin{split}
\LL_Y^t & =
  - \, M_T\,\bar{T}_R T_L \; - \, m_t\,\bar{t}_R t_L
  \\ & \quad
  + \, s c \frac{m_t}{F}        \, H \bar{T}_R T_L \;
  - \, \frac{s}{c}\frac{m_t}{v} \, H \bar{T}_R t_L \;
  - \, \frac{m_t}{v}            \, H \bar{t}_R t_L
  \\ & \quad
  + \, \frac{i}{cs}\frac{m_t}{v}\beta \, \eta\,\bar{T}_R T_L \;
  + \, i\frac{m_t}{v}\beta'           \, \eta\,\bar{t}_R T_L \;
  - \, i \frac{m_t}{F}\beta''         \, \eta\,\bar{t}_R t_L \;
  + \, \cdots \, + \, \hc
\end{split}
\end{align}
The coefficients of the axion couplings depend on the fermion
$U(1)_\eta$ charges as:
\begin{subequations}
\begin{align}\label{beta}
  \beta &=\; \frac{1}{\sqrt5}\left[(2-\beta_0-\beta_1)s^2 - \beta_2
  c^2\right], \\
  \beta' &=\; \frac{1}{\sqrt5}\left(2 - \beta_0 - \beta_1 +
  \beta_2\right), \\ 
  \beta'' &=\; \frac{1}{\sqrt{5}}\left[(2 - \beta_0 + \beta_2)s^2 +
  \beta_1 c^2\right]
\end{align}
\end{subequations}

From the mass-eigenstate Lagrangian of Eq.~(\ref{yuk-exp2}) we can
read off the vertex structure.  By construction, $\eta$ is CP-odd.
While the particular values of the Yukawa coupling coefficients are
specific to a given model, their orders of magnitude with respect to
the mass scales of the theory, shown in Table~\ref{tab:vertices}, are
dictated by the $SU(2)$ transformation properties and thus are generic
for Little Higgs models.  The relation of the $T\bar{T}$ and
$t\bar{t}$ couplings is a consequence of the fact that, after Little
Higgs symmetry breaking, the heavy states are vectorlike, while $t$ is
a chiral fermion.  For $T\bar{T}$, unbroken EW symmetry forbids a
coupling to the Higgs doublet, so this coupling must be proportional
to~$v/F$ which arises from $t-T$ mixing.  Similarly, the coupling
$t\bar{t}\eta$ is forbidden for unbroken symmetry since $\eta$ is a
singlet.  The chirality assignments of the mixed couplings are
model-dependent: If the heavy singlet $T$ is replaced by a doublet,
$P_R$ and $P_L$ have to be exchanged.  For $T$ as a heavy triplet, the
structure of $T\bar{t}H$ is unchanged while the $T\bar{t}\eta$
couplings are suppressed by $v/F$.

\begin{table}[htb]
\renewcommand{\arraystretch}{1.2}
\begin{displaymath}
  \begin{array}{|cc|}
    \hline
    \bar{T}TH & {\cal O}(\frac{v}{F}) \\
    \bar{T}tH & {\cal O}(1)\,\pl + {\cal O}(\frac{v}{F})\,\pr\\
    \bar{t}tH & {\cal O}(1)\\
    \hline
  \end{array}
  \qquad
  \begin{array}{|cc|}
    \hline
    \bar{T}T\eta & {\cal O}(1)\,\gamma_5 \\
    \bar{T}t\eta & {\cal O}(1)\,\pr + {\cal O}(\frac{v}{F})\,\pl \\
    \bar{t}t\eta & {\cal O}(\frac{v}{F})\, \gamma_5\\
    \hline
  \end{array}
\end{displaymath}
\caption{Chiral structure of the Yukawa couplings in Little Higgs
  models with a heavy $SU(2)$ singlet $T$.  $P_{R,L}$ are the
  chirality projectors.  For the coefficients, only the order of
  magnitude is indicated. This structure remains valid for the $\mu$
  model presented in Sec.~\ref{sec:mumodel}.}
\label{tab:vertices}
\end{table}

Although not previously defined in the literature, we posit that the
Yukawa coupling for down-type fermions in the Littlest Higgs model can
be written down as:
\begin{equation}
\LL_Y^b = 
- \frac{\lambda_b}{\sqrt{2}} F \xi^{\beta_3} \overline{b}_R \Tr 
  \left[ \Sigma Y_1 \chi_L \right] + \hc \, ,
\end{equation}
where $Y_1$ is one of the hypercharge operators defined
in~\cite{KR03},
\bq
Y_1 = \frac{1}{10} \text{diag}( 3, 3, -2, -2, -2) \, .
\eq
The pseudoscalar coupling is then
\bq
g_{\eta bb} \overline{b} \gamma^5 b \qquad\text{with} \qquad 
g_{\eta bb} = \frac{m_b}{\sqrt{5} F} \beta_3
\eq
Clearly, $\beta_3$ is an additional new free parameter.

The free parameters in the model are then one Yukawa coupling,
$\lambda_1$, the mass scale $F$, and the four differences of $U(1)$
fermion charge assignments, $\beta_{0,1,2,3}$; $\lambda_2$ is fixed by
the top quark mass condition.  We consider the case $F=4$~TeV, which
is motivated by the precision EW limits for the Littlest
model~\cite{KR03}.

\begin{figure}[htb]\begin{center}
\includegraphics[scale=0.9]{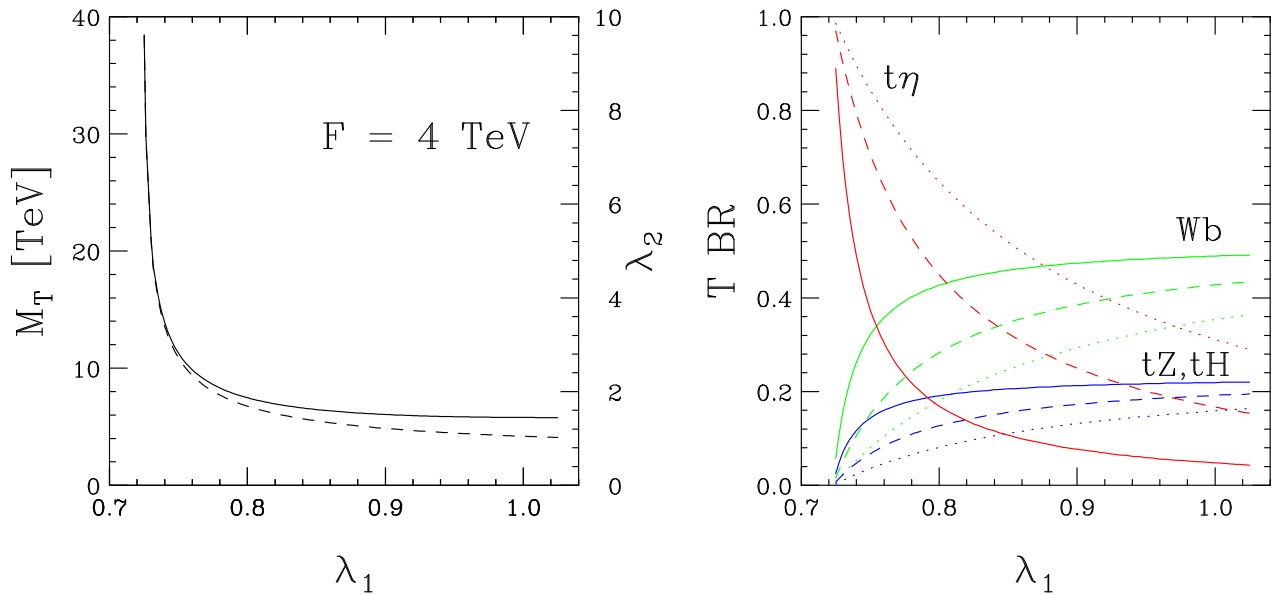}
\vspace{-7mm}
\caption{Left: $M_T$ (solid) v. $\lambda_1$ in the Littlest Higgs model 
with $F=4\,\TeV$. Also plotted is $\lambda_2$ (dashed), which is
constrained by the choice of $\lambda_1$ and the known value $m_t$.
Right: Heavy $T$ quark branching ratios to $Wb$ (green), $tZ/th$
(blue), and $t\eta$ (red), for three choices of the $U(1)_\eta$ charge
differences $\beta_{0,1,2}$: 1,1,1 (solid); 1,0,1 (dashed); 0,0,1
(dotted).}
\label{fig:LH.TBR}
\vspace{-3mm}
\end{center}\end{figure}

The presence of the pseudo-axion $\eta$ alters the decay spectra of
the heavy quark $T$ (see Appendix~\ref{sec:decay}), drastically for
some regions of parameter space.  Note in Eq.~(\ref{yuk-exp2}) that
the $T\to t\eta$ partial width is proportional to $\beta'^2$, which is
a pure difference of the $\beta_i$ and contains no mixing factors,
whereas the $T\to tH$ partial width is protected by $SU(2)$ and
therefore appears only as a consequence of $T-t$ mixing.  The latter
is true also of the decays $T\to Wb,tZ\,$.  (Ignoring mass effects,
$\Gamma_{tH}=\Gamma_{tZ}=\frac{1}{2}\Gamma_{bW}\,$.)  Due to the sum
rule $1/\lambda^2_t = 1/\lambda^2_1 + 1/\lambda^2_2$, both $\lambda_1$
and $\lambda_2$ must be larger than $\lambda_t$.  While there is no
general obstacle against $\lambda_1>\lambda_2$ (i.e. $\sin>\cos$),
this is disfavored by large mixings of the third-generation EW
currents~\cite{KR03}.  So we take $\lambda_1$ to be bounded from below
by $\lambda_t$, and from above by $\sqrt{2}\lambda_t$ using the
condition $\lambda_1=\lambda_2$.  This translates to a range
$0.72<\lambda_1<1.02$.  In Fig.~\ref{fig:LH.TBR} we plot $m_T$ v. the
allowed values of $\lambda_1$, as well as the consequential
$\lambda_2$, which closely tracks $m_T$.  As expected, in the
$\lambda_1\to\lambda^{SM}_t$ limit, $\lambda_2$ becomes large and
almost totally responsible for $m_T$, which becomes (perhaps
unnaturally, although one could argue that $\lambda_1=\lambda^{SM}_t$
is ``natural'' as well) much larger than $F$.  For the lower extremal
limit, $t-T$ mixing vanishes and the partial decay widths to
$Wb,tZ,tH$ vanish correspondingly (the decoupling limit).  The
right-hand side of Fig.~\ref{fig:LH.TBR} shows the resulting $T$ BRs
for three choices of the relevant $\beta_i$.  For large mixing and
equal $U(1)_\eta$ charges, BR($t\eta$) bottoms out at around $5\%$ and
dominates only when the mixing becomes very small and $m_T$ grows
quite large.  For other choices of $\beta_i$, e.g. only $\beta_2\ne
0$, $\beta'$ triples and BR($t\eta$) dominates everywhere, with the
$T$ quark total width being ${\cal O}(10-100)$~GeV.  For negative
$\beta_{0,1}$, which are in no way unnatural, the non-$\eta$ decays
would be practically squelched for all Yukawa couplings, and the total
$T$ width can be hundreds of GeV.  The shapes of all curves in
Fig.~\ref{fig:LH.TBR} are independent of the choice of $F$.

The dominant decays of the $\eta$ are to $b\bar{b}$, $gg$ and
$\gamma\gamma$, the latter two being loop-induced couplings (see
Appendix~\ref{sec:loop}), and for $m_\eta>2m_t$, decays to top quark
pairs.  While there will also be loop-induced decays to $ZZ$ and
$W^+W^-$, the BRs to observable final states are small fractions of
the already rare decay rates, so we do not consider these further.
There are in principle also direct couplings to the lighter fermions,
but proportional to the fermion mass squared.  We ignore these, as the
only straightforward, high-efficiency signal would be to muons and
that BR is typically an order of magnitude smaller than to
photons. The BRs are independent of the scale $F$, since it appears
the same way in all partial widths; and to the percent level,
independent of $\lambda_1$.  Table~\ref{tab:L-BR} shows the BRs to the
dominant final states for various values of $m_\eta$.  Note that from
$m_H$ stability arguments we don't expect $m_\eta\gtrsim v$, but this
is order-of-magnitude, so we show one case with $m_\eta>2m_t$ for
illustration.

\begin{table}[h]
\begin{center}
\begin{tabular}{|l@{\hspace{0.5ex}}|r|r|r|r|r|r|}
\hline
\multicolumn{1}{|c|}{$m_\eta$ [GeV] ($\beta_i$)} &
\multicolumn{1}{c|}{150 (1,1,1,1)} &
\multicolumn{1}{c|}{150 (1,0,1,1)} &
\multicolumn{1}{c|}{150 (1,1,1,0)} &
\multicolumn{1}{c|}{250 (1,1,1,1)} &
\multicolumn{1}{c|}{400} \\ \hline
$\eta\to t\bar{t}$
  & $ 0\%$   & $ 0\%$   & $ 0\%$   & $ 0\%$   & $99.3\%$  \\ \hline
$\eta\to b\bar{b}$
  & $38\%$   & $72\%$   & $ 0\%$   & $17\%$   & $0.02\%$  \\ \hline
$\eta\to gg$
  & $61\%$   & $29\%$   & $99.7\%$ & $83\%$   & $0.5\%$   \\ \hline
$\eta\to \gamma\gamma$
  & $0.17\%$ & $0.08\%$ & $0.27\%$ & $0.27\%$ & $0.002\%$ \\ \hline
\end{tabular}
\caption{Dominant branching ratios of the $\eta$ in the Littlest Higgs
model.  The case $\beta_i=(0,0,1,1)$ is identical to (1,0,1,1).  For
the case $m_\eta=400$~GeV, the $\beta_i$ assignments are irrelevant,
as decay to top quark pairs overwhelmingly dominates.}\label{tab:L-BR}
\end{center}
\end{table}

It should be obvious that the phenomenology of the Littlest Higgs
model has an extreme dependence on the presence of the pseudo-axion
and the pattern of $U(1)_\eta$ fermion charge assignments $\beta_i$,
which are undefined.  It would not be inaccurate to say that the model
is fairly incompletely defined.  The situation is even more
complicated than presented above, since there should be some
$U(1)_\eta$ charge assignment for {\it every} SM fermion, greatly
increasing the number of free parameters, although one could
reasonably argue based on flavor-changing neutral current (FCNC)
constraints that the charge assignments are identical at least across
generations.  Even the presence of just the third-generation $\beta_i$
introduces enough new parameters to present a serious phenomenological
challenge at future colliders.  The positive aspect of this is that
measuring these parameters may provide clues to the model's UV
completion.  Furthermore, the Little Higgs mechanism doesn't depend
on any of the $\beta_i$.


\subsection{The {\boldmath $\mu$} model}
\label{sec:mumodel}

Recently, Schmaltz has proposed a moose-type model~\cite{simplest}
where the Little Higgs mechanism is implemented in a very economic
way.  A $\mu$ term which explictly breaks some of the global symmetry
is responsible both for the absence of fine-tuning and for EWSB.  The
EW group is enlarged to gauged $SU(3)
\times U(1)$.  There are two nonlinear sigma model fields,
each of which parameterizes a coset space $U(3)/U(2)$:
\bq
\Phi_1 = \exp\biggl[  i \frac{F_2}{F_1} \Theta\biggr]
\begin{pmatrix} 0 \\ 0 \\ F_1 \end{pmatrix} \, , 
\qquad
\Phi_2 = \exp\biggl[ -i \frac{F_1}{F_2} \Theta\biggr]
\begin{pmatrix} 0 \\ 0 \\ F_2 \end{pmatrix} \, , 
\eq
where~\footnote{There are other possible choices for the generator
$T_\eta$ that multiplies the $\eta$ field, e.g., $T^8$ or
$\mathrm{diag}(0,0,1)$.  However, after EWSB these choices introduce
kinetic mixing of the $\eta$ with unphysical Goldstone bosons.
Removing this mixing by appropriate field redefinitions is equivalent
to choosing $T_\eta$ proportional to the unit matrix.}
\bq
\Theta = \frac{1}{F} \left\{ \frac{\eta}{\sqrt{2}} +
                             \begin{pmatrix}
                               \begin{array}{cc}
                                 0 & 0 \\ 0 & 0
                               \end{array} & h^* \\
                               h^T & 0
                             \end{pmatrix} \right\} \, ,
\qquad
F^2 = F_1^2 + F_2^2 \; .
\eq
As usual, the global Little Higgs symmetry is broken radiatively by
the gauge and Yukawa interactions that trigger EWSB and give the Higgs
a mass, but they leave a global $U(1)_\eta$ symmetry intact.  The
Coleman-Weinberg mechanism thus generates a negative Higgs
mass-squared and a positive quartic coupling, but without $U(1)_\eta$
breaking it does not contribute $\eta^2$, $\eta^4$, or $\eta^2 h^2$
terms.  These are generated only by the $\mu$ term:
\bq
  -V = \; \mu^2 \Phi_1^\dagger \Phi_2 + \text{h.c.} = 2 F_1 F_2 \mu^2
  \cos\left( \frac{F \eta}{\sqrt{2} F_1 F_2} \right) \Biggl[
  1 - \frac{F^2}{2 F_1^2 F_2^2} (h^\dagger h) +
  \frac{F^4}{24 F_1^3 F_2^3} (h^\dagger h)^2 + \ldots \Biggr]\,.
\eq

Thus, the complete potential up to quartic order is ($\kappa\equiv
F_1/F_2 + F_2/F_1 \geq 2$):
\bq\label{eq:mumodelpot}
-V = \;
- (\delta m^2 + \mu^2 \kappa ) (h^\dagger h)
- \mu^2 \kappa \frac{\eta^2}{2}
+ \left( \frac{\mu^2 \kappa^2}{12 F_1 F_2} - \delta\lambda \right) 
  (h^\dagger h)^2
+ \frac{\mu^2\kappa^2}{12 F_1 F_2}
  \left( \frac{\eta^4}{4} + \frac{3 (h^\dagger h)\eta^2}{2} \right)
+ \ldots
\eq
Here, $\delta m$ and $\delta\lambda$ are the one-loop contributions to
the Higgs boson mass and quartic coupling from the Coleman-Weinberg
potential given in~\cite{simplest}. From this we read off the $\eta$
mass
\bq
m_\eta \; = \; \sqrt{\kappa}\,\mu \; \geq \; \sqrt{2}\,\mu.
\eq
To minimize the amount of fine-tuning, $\mu$ is chosen to be roughly
of the order of the EW scale. Also from Eq.~(\ref{eq:mumodelpot}) we
deduce the connection
\bq\label{eq:sumrule}
m_H^2 = -2(\delta m^2 + m_\eta^2) \; .
\eq
The interesting property of this model is that the $\eta$ mass is
predicted, and can be calculated based on other experimental
observables.  In principle, only the masses of the Higgs boson and
either the heavy $T$ quark or one heavy vector are needed to determine
the scales $F_1$ and $F_2$.  One can then calculate $\delta
m^2$~\cite{simplest} and predict $m_\eta$ by Eq.~(\ref{eq:sumrule}).
However, $\delta m^2$ depends on the theory cutoff $\Lambda$ and has a
noticeable dependence, so in practice performing precision fits to the
model would be difficult, since the physical impact of the cutoff
would not necessarily be clear.  Nevertheless, within some reasonable
uncertainty, measuring these values would be a first test of whether
or not a discovery satisfied this particular model.

There are two possible gauge charge assignments for fermions in the
$\mu$ model, called types I and II.  The first has heavy partners of
the top, charm and up quarks, and all generations carry identical
gauge quantum numbers. This is, however, anomalous.  This model also
appears to be ruled out from precision EW data~\cite{simplest}, so we
do not discuss it further.  The second model has slightly different
charge assignments to eliminate the anomalies, and is not yet ruled
out.  In this scenario the top, strange and down quarks have heavy
partners, and the Yukawa interactions are given by
\begin{align}
\begin{split}
\LL & = 
- \, \lambda^t_1 \overline{t}_{1,R} \Phi_1^\dagger \Psi_{T,L} \,
- \, \lambda^t_2 \overline{t}_{2,R} \Phi_2^\dagger \Psi_{T,L} \,
- \, \frac{\lambda^b}{\Lambda} \epsilon^{ijk} \Phi^i_1 \Phi^j_2 \Psi^k_{T,L} \,
\\ & \quad
- \, \lambda^{d}_1 \overline{q}^{d}_{1,R} \Phi_1^\dagger \Psi_{Q,L} \,
- \, \lambda^{d}_2 \overline{q}^{d}_{2,R} \Phi_2^\dagger \Psi_{Q,L} \,
- \, \frac{\lambda^u}{\Lambda} \epsilon^{ijk} \Phi^i_1 \Phi^j_2 \Psi^k_{Q,L} \,
\\ & \quad
+ \cdots + \hc 
\end{split}
\end{align}
where the $SU(3)$ triplet is $\Psi_{T,L} = (\tau^2 q^3_L, T_L)^T$ and
$d,u$ sum over the first two generations; we assume the up-quark
sector to be generationally diagonal, which is not necessary but
simplifies the model.  We diagonalize the mass matrix first with
${\cal O}(1)$ right-handed singlet-field rotations $t_{1,R}\to
c\,t_R+s\,T_R$, $t_{2,R}\to -s\,t_R+c\,T_R$, with $s=\lambda_1
F_1/M_T$ and $c=\lambda_2F_2/M_T$.  We then rotate the left-handed
fields by an ${\cal O}(v/F)$ angle $N_1 m_t/M_T$ (see below).  The
leading (to ${\cal O}(v/F)$) Yukawa terms are
\begin{align}\label{muyuk}
\begin{split}
\LL & = - M_T \,\overline{T}_R T_L - m_t \,\overline{t}_R t_L
\\ & \quad
        - \frac{m_t}{v} \,  H \overline{t}_R t_L
        - g_{\eta TT}   \,  H \overline{T}_R t_L
        + g_{\eta tt}   \,  H \overline{t}_R T_L
        + g_{HTT}       \,  H \overline{T}_L T_R
\\ & \quad
        + i g_{\eta tt}   \, \eta\,\overline{t}_R t_L
        + i g_{H TT}      \, \eta\,\overline{T}_R t_L
        + i \frac{m_t}{v} \, \eta\,\overline{t}_R T_L
        + i g_{\eta TT}   \, \eta\,\overline{T}_R T_L
\\ & \quad
        + \cdots + \hc 
\end{split}
\end{align}
where $m_t=\lambda_1\lambda_2 vF/\sqrt{2}M_T$ and $M_T=
\sqrt{\lambda_1^2 F_1^2 + \lambda_2^2 F_2^2}\,$.  The couplings are
given by
\bq
g_{\eta tt} = \frac{m_t N_2}{\sqrt{2} F} - \frac{m_t^2 N_1}{v M_T}
\, , \qquad
g_{\eta TT} = \frac{N_1 m_t}{v}
\, , \qquad
g_{HTT} = - \frac{N_1^2 m_t^2}{v  M_T} + \frac{N_3 v}{2 M_T} \; ,
\eq
with the abbreviations 
\bq
N_1 = \frac{F_1F_2}{F^2}\frac{\lambda_1^2-\lambda^2_2}{\lambda_1\lambda_2}
\, , \qquad
N_2 = \frac{F_2^2 - F_1^2}{F_1F_2}
\, , \qquad
N_3 = \frac{\lambda_1^2 F_2^2 + \lambda_2^2 F_1^2}{F^2} \; .
\eq

Comparing this with the Littlest Higgs model, we observe that here the
axion properties are fully defined, i.e., no additional parameters
have to be introduced.  Furthermore, there are heavy partners of at
least one quark and lepton of each generation.  By contrast, in the
Littlest Higgs model the presence of heavy fermions beyond the
top-quark partner $T$ is not necessary.

We adopt the parameter set of Ref.~\cite{simplest}, consistent with
existing EW and flavor data and the preference for a light Higgs
boson.  The complete set of inputs, excluding the lepton sector, is
$F_{1,2}$, $\lambda^i_{1,2}$, $\lambda^b$, $\lambda^u$, $\lambda^d$,
and $\mu$.  To simplify the set we assume
$\lambda^{1,2}_1\ll\lambda^{1,2}_2\equiv\lambda^t_2$, so there is
essentially no mixing between the SM down-type quarks and their heavy
partners, whose masses are simply $\lambda_2 F_2$.  Deviations from
this matter very little for the gross phenomenology.  We further
choose $\lambda^t_{1,2}$ to minimize $m_T$ given $F_{1,2}\,$ and
$m_t$.  $\lambda^b$ and $\lambda^u$ are fixed by $m_b$, $m_c$ and
$m_u$.  Our only free parameters are then $F_{1,2}$ and $\mu$, with
the constraint that $F_1$ not be as small as the EW scale, $F\gtrsim
2$~TeV from EW precision constraints (primarily $\Delta T$ and
four-fermion operators), and $F_2>F_1$ to avoid too much mixing which
would lead to fermion non-universality.  $F_1=F_2$ lies right at the
edge of the limits on the latter.  The cutoffs of the theory,
$\Lambda_{1,2}$, are nominally $4\pi F_{1,2}$, but this is somewhat
vague.  Their precise choice affects the predicted Higgs mass, but
because of the inherent uncertainty as to their definition we simply
choose the suggested~\cite{simplest} value of
$\Lambda_1=\Lambda_2\equiv\Lambda=5$~TeV and vary $\mu$ over a
slightly broader range than the calculated $m_H$ would suggest is
allowed or favored by data.

Starting then with the ``Golden Point'' in Ref.~\cite{simplest} of
\bq
F_1 = 0.5 \,\TeV, \quad F_2 = 2 \,\TeV, \quad \Lambda = 5 \,\TeV
\eq
we calculate 
\bq
m_T = 990\,\GeV, \quad m_{S,D} = 700\,\GeV, \quad
m_{Z'} = 1.2\,\TeV, \quad m_{W'} = 950\,\GeV \, .
\eq
For $\mu\gtrsim 150$~GeV, $m_H$ is lower than the LEP direct exclusion
limit of $\sim 114$~GeV.  For $\mu\lesssim 120$~GeV we find that $m_H$
is greater than the SM $95\%\;{\rm c.l.}$ upper bound from precision
EW data.  However, in general in Little Higgs models, larger values of
$m_H$ are allowed because of additional positive $\Delta T$
contributions from the other new content.  We therefore do not
restrict ourselves to a lower bound on $\mu$.  For this Golden Point,
as $\mu\to 0$, $m_H$ plateaus at around 450~GeV.  To illustrate the
dependence on $\Lambda$, if instead we explicitly set $\Lambda=4\pi
F_1$ then only $\mu\gtrsim 135$~GeV is allowed.  It is noteworthy that
$m_\eta<2m_t$ for this parameter choice.

\begin{figure}[h]\begin{center}
\includegraphics[scale=0.9]{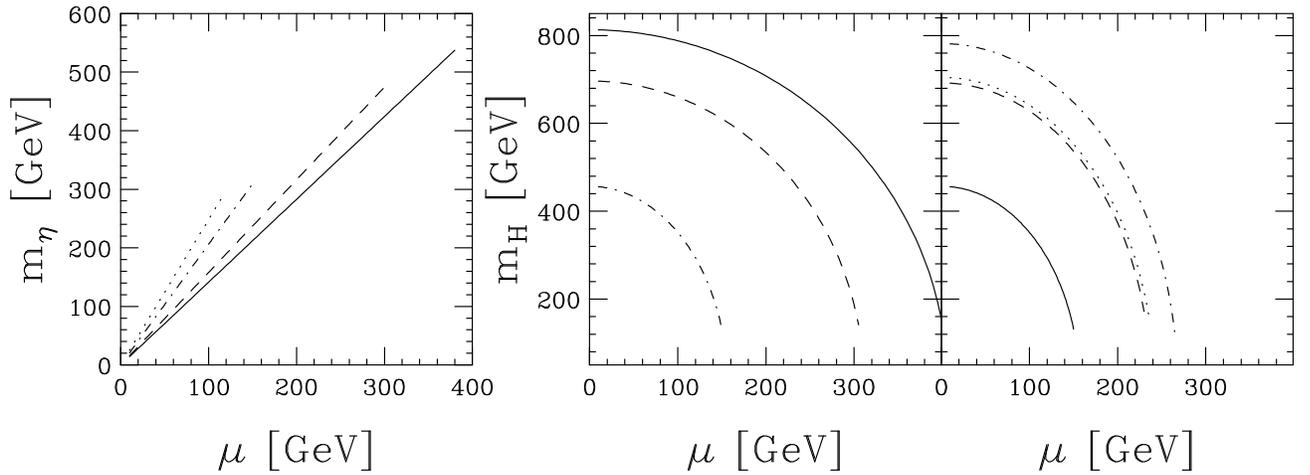}
\vspace{-7mm}
\caption{Left: $m_\eta$ v. $\mu$ for four choices of the ratio $F_1/F_2$: 
1 (solid), 1/2 (dashed), 1/4 (dot-dashed), and 1/6 (dotted).  Middle:
$m_H$ v. $\mu$ for fixed $F_2=2.0$~TeV and various choices of
$F_1/F_2$: 1 (solid), 1/2 (dashed) and 1/4 (dot-dashed).  Right: $m_H$
v. $\mu$ for fixed $F_1/F_2=1/4$ and various $F_1$ [TeV]: 0.5 (solid),
1.0 (dashed), 1.5 (dot-dashed), and 2.0 (dotted).}
\label{fig:mu.mass}
\vspace{-5mm}
\end{center}\end{figure}
\begin{figure}[htb]\begin{center}
\includegraphics[scale=0.9]{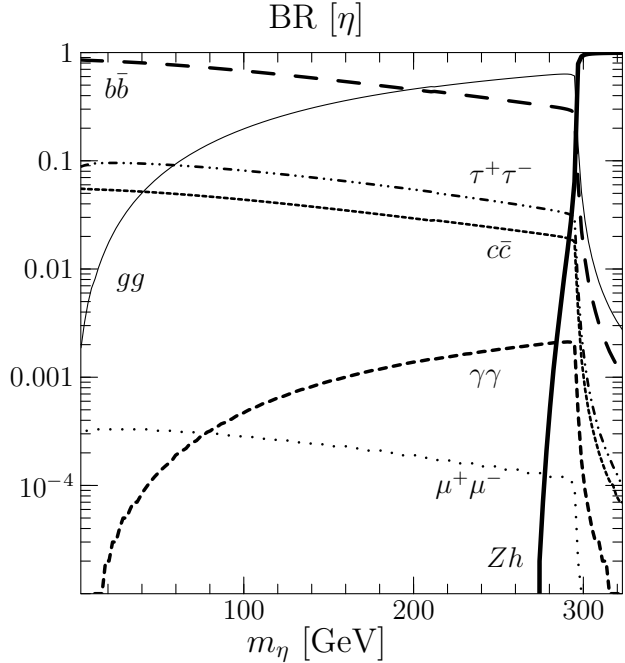}
\caption{$\eta$ branching ratios in the $\mu$ model for the Golden 
Point, as discussed in the text, as a function of $m_\eta$.  }
\label{fig:mu.etaBR}
\vspace{-5mm}
\end{center}\end{figure}

Moving away from the Golden Point, we observe several interesting
features.  First, while $m_\eta$ depends only on the ratio $F_1/F_2$
rather than their individual magnitudes, and grows larger for smaller
ratios, for smaller ratios it is quite easy to violate the LEP direct
exclusion limit on $m_H$.  The trend for increasing ratio is shown in
Fig.~\ref{fig:mu.mass}, left panel.  The upper end of each curve is
cut off by the condition that $m_H\gtrsim 114$~GeV.  In general as
$F_1/F_2$ approaches about 1/3, $m_\eta>2m_t$ becomes possible for
$\mu\gtrsim 200$~GeV.

The right panel shows the (more complicated) trend for $m_H$.  As
$F_1$ moves closer to $F_2$, $m_H$ prefers to become quite heavy,
although as both values become very large, this trend can reverse and
$m_H$ becomes lighter.  For example, $m_H$ plateaus at around 800~GeV
for small $\mu$, although at the extremal point $F_i=3.3$~GeV the
range of $m_H$ suddenly drops radically to around 100-200~GeV.  For
$F_i\gtrsim 3.4$~TeV, EWSB does not occur, providing a strong upper
bound on the allowed values of $F_i$.  For smaller ratios $F_1/F_2$,
both the low-$\mu$ plateau for $m_H$ and the upper limit on $\mu$
become smaller.  The latter is cut off by the Higgs becoming lighter
than the LEP direct exclusion limit.  A more thorough understanding
of the allowed parameter space based on EW precision data would
require a complete EW fit, which is beyond the scope of this work.  We
would still expect the EW data to prefer a light Higgs, but the upper
limit could be considerably larger than in the SM, which is why we do
not impose an upper limit on $m_H$ in Fig.~\ref{fig:mu.mass}.

Decays of the $\eta$ are more straightforward in this model than that
of Sec~\ref{sec:littlest}, because all couplings are precisely defined.
This is true for the $b\bar{b}\eta$ coupling as well, which in this
model we derive to be
\bq
g_{\eta bb} \, = \, - \frac{N_2}{\sqrt{2}}\frac{m_b}{F} \; ,
\eq
which is exactly like the $t\bar{t}\eta$ coupling (Eq.~(\ref{muyuk})),
except that there is no term proportional to $N_1$ from mixing since
there is no heavy partner of the $b$ quark.  It has the interesting
properties that it appears only at higher orders in the expansion of
$\Phi_{1,2}$ (yet remains ${\cal O}(v/F)$), and that it is
driven rapidly to zero as $F_1$ and $F_2$ approach each other.
Couplings to muons and tau leptons will have the same form.  The
coupling to charm quarks will technically have the same form as that
for top quarks, but here the mixing term is trivially small due to the
ratio $m_c/M_C$.  We can then straightforwardly predict the partial
widths for $\eta$ to decay into SM particles.

However, there is one complication: unlike in the Littlest model, here
there is a $ZH\eta$ coupling which arises from the kinetic term for
$\Phi_1,\Phi_2$ (note that both Higgs triplets have the same $SU(3)_w
\times U(1)_x$ quantum numbers).
\bq\label{eq:kinmumodel}
\LL = (D_\mu \Phi_1)^\dagger D^\mu \Phi_1 + (D_\mu \Phi_2)^\dagger
    D^\mu \Phi_2
\eq
with the covariant derivative expressed by the physical fields:
\bq
D_\mu \Phi_k = \partial_\mu \Phi_k + i
\begin{pmatrix}
  e A_\mu + g \frac{c_w}{2} (1-t_w^2) Z_\mu & \frac{g}{\sqrt{2}}
  W_\mu^- & 0 \\ \frac{g}{\sqrt{2}} W_\mu^+ & - \frac{g}{2 c_w}
  Z_\mu & 0 \\ 0 & 0 & 0
\end{pmatrix} \Phi_k + \ldots \qquad k = 1,2 \; .
\eq
The dots stand for the part of the covariant derivative containing the
five remaining heavy gauge bosons, which we leave out here.  Plugging
in the expansion of the nonlinear sigma-model Higgs fields, we derive
a $ZH\eta$ coupling
\bq
\LL_{ZH\eta} = \frac{m_Z}{\sqrt{2} F} N_2 Z_\mu
                    (\eta \partial^\mu H - H \partial^\mu \eta)
\eq
with $N_2$ defined previously.  Such a coupling is forbidden for
unbroken EW symmetry ($\eta$ is a singlet) so it must be
proportional to $v/F$, but this is partially compensated for by large
$N_2$ for the Golden Point.

We plot the BRs for $\eta$ to decay into SM particles, including the
$ZH$ final state, in Fig.~\ref{fig:mu.etaBR}, as a function of
$m_\eta$ for the Golden Point.  Decays to the final state $b\bar{b}$
dominate for $m_\eta\lesssim 200$~GeV.  Above this $gg$ dominates
until the $ZH$ threshold is crossed at around $m_\eta=290$~GeV.  Note
that BR($\gamma\gamma$) remains non-trivial over an extremely large
range of $\mu$.  We can make a rough comparison with how a typical
two-Higgs doublet model pseudoscalar behaves at low values of
$\tan\beta$.  Here, the role of $\tan\beta$ is played by the ratio of
$\Phi$ decay constants, $F_1/F_2$.  Due to constraint from lepton
universality, we must have $F_1<F_2$, or $\tan\beta<1$, which is the
opposite of what is required in e.g. the MSSM.

Although we do not plot the decay BRs for the Higgs, it is worth
pointing out that for much of the parameter space where the $\eta$
does not decay to $ZH$, instead the decay $H\to Z\eta$ is
kinematically allowed and constitutes a very large partial width.  We
discuss the impact of this on general Higgs sector phenomenology
further in Sec.~\ref{sec:pheno}.


\subsection{The original Simple Group model}
\label{sec:simplest}

The Simple Group model introduced in Ref.~\cite{simple} displays an
axion phenomenology that is a synthesis of the two previously
discussed models.  The low-energy effective theory contains two Higgs
doublets.  The global symmetry structure exhibits breaking of the
semisimple group $[SU(4)]^4 \to [SU(3)]^4$, while one $SU(4)$ subgroup
is gauged and broken down to the SM gauge group $SU(2)$. Neglecting
heavy singlet scalars, the four fundamental scalar multiplets can be
approximately written as
\begin{align}
  \Phi_1 &\approx\; e^{+\ii H/F} e^{+\ii E_u / F} (0,0,F,0)^T \, , \\
  \Phi_2 &\approx\; e^{-\ii H/F} e^{-\ii E_u / F} (0,0,F,0)^T \, , \\
  \Psi_1 &\approx\; e^{+\ii H/F} e^{+\ii E_d / F} (0,0,0,F)^T \, , \\
  \Psi_2 &\approx\; e^{-\ii H/F} e^{-\ii E_d / F} (0,0,0,F)^T \, .
\end{align}
The matrices are
\bq
H = \frac{1}{\sqrt{2}}
  \begin{pmatrix}
    \begin{matrix}
      0 & 0 \\ 0 & 0
    \end{matrix} &
    \begin{matrix}
      h_u & h_d
    \end{matrix} \\
    \begin{matrix}
      h_u^\dagger \\ h_d^\dagger
    \end{matrix} &
    \begin{matrix}
      0 & 0 \\ 0 & 0
    \end{matrix}   
  \end{pmatrix}
\, , \qquad
E_U = \frac{\eta_u}{6} \text{diag} (1,1,-3,1)
\, , \quad
E_D = \frac{\eta_d}{6} \text{diag} (1,1,1,-3)
\; .
\eq
We introduce the abbreviations
\bq
\xi_{u/d} = \exp[\ii \eta_{u/d}/(6F)]
\eq
to obtain
\begin{eqnarray}
  (\Phi_1)_i &\approx\; \xi_u^3 &\left[ F \delta_{i,3} +
  \frac{\ii}{\sqrt{2}} h_{u,j} \delta_{ij} \right] , \\
  (\Phi_2)_i &\approx\; \xi_u^{-3} &\left[ F \delta_{i,3} -
  \frac{\ii}{\sqrt{2}} h_{u,j} \delta_{ij} \right] , \\
  (\Psi_1)_i &\approx\; \xi_d^3 &\left[ F \delta_{i,4} +
  \frac{\ii}{\sqrt{2}} h_{d,j} \delta_{ij} \right] , \\
  (\Psi_2)_i &\approx\; \xi_d^{-3} &\left[ F \delta_{i,4} -
  \frac{\ii}{\sqrt{2}} h_{d,j} \delta_{ij} \right] .
\end{eqnarray}

Considering the potential necessary to break EW symmetry, we can
construct four bilinears $\Phi_1^\dagger \Psi_1$,
$\Phi_2^\dagger\Psi_2$, $\Phi_1^\dagger \Psi_2$, and $\Phi_2^\dagger
\Psi_1$. In those with like indices the Higgs doublets cancel, so
these terms approximately vanish.  The potential for $\eta$ is
generated by the remaining terms
\begin{align}
V_\eta & = \; 2 b_{12} \Re \left[ \Phi_1^\dagger \Psi_2 \right]
            + 2 b_{21} \Re \left[ \Phi_2^\dagger \Psi_1 \right] \, ,
\end{align}
namely,
\bq
\Phi^\dagger_1 \Psi_2 \approx
F^2 (\xi_u \xi_d)^3 \left( \exp[-2\ii H/F] \right)_{34} 
= - (\xi_u \xi_d)^3 (h_u^\dagger h_d)_{34}
\; , \qquad
\Phi^\dagger_2 \Psi_1 \approx (\Phi^\dagger_1 \Psi_2)^\dagger \; .
\eq
The coefficients $b_{ij}$ are expected to be of order $F^2$.  The
expansion in powers of $1/F$ yields
\bq
  V_\eta = 2 (b_{12} + b_{21}) \biggl[ \Re \left[ h_u^\dagger h_d
  \right] - \frac{\eta_u + \eta_d}{2 F} \Im \left[ h_u^\dagger h_d
  \right] - \frac{(\eta_u + \eta_d)^2}{8 F^2} \Re \left[ h_u^\dagger h_d
  \right] \biggr].
\eq
Note that the $\eta$ couplings prefactor cannot vanish if EW symmetry
is to be broken.  Introducing the standard Higgs-field components
(cf.~also \cite{Csaki2}), we get (as usual in 2-Higgs-doublet models,
the scalar Higgs bosons are denoted by $h$ and $H$, the pseudoscalar
by $A$)
\begin{align}
\Re \left[ h_u^\dagger h_d \right] & = \;
\frac12 \left[ 
  (v_1 + H c_\alpha - h s_\alpha)( v_2 + H s_\alpha + h c_\alpha)
       + c_\beta s_\beta (A^2 + 2 H^+ H^-)
\right] \\
\Im \left[ h_u^\dagger h_d \right] & = \;
\frac{A}{2} \left[
  v_1 c_\beta - v_2 s_\beta + H c_{\alpha+\beta} - h s_{\alpha-\beta}
\right]
\end{align}
We observe that a potential is generated only for the linear
combination $\eta_+ = (\eta_u + \eta_d)/\sqrt{2}$.  This pseudo-axion
is analogous to the one in the $\mu$ model.  However, the mass of
$\eta_+$ is very low, of order $v^2/F$.  A further effect of the above
potential is mixing of $\eta_+$ with the pseudoscalar $A$ of the
second Higgs doublet, suppressed by $v/F$.

The orthogonal combination $\eta_- = (\eta_u - \eta_d)/\sqrt{2}$
remains massless, similar to the $\eta$ of the Littlest Higgs model.
Without introducing extra symmetry-breaking terms we have no
prediction for $m_{\eta_-}$.

Due to the mixing interaction, there is an $h\eta_+\eta_+$ vertex with
a coupling of the order $v^3/F^2$.  This modifies the standard Higgs
width by an amount of
\bq
\Gamma_{H\to\eta_+\eta_+} \sim
\frac{1}{16\pi} \sqrt{1-\frac{4 m_{\eta_+}^2}{m_H^2}}
\frac{v^5}{F^4} \sim
\frac{15}{\left(F[\text{TeV}]\right)^4} \; \MeV,
\eq
and gives a BR of order $5-10\%$ into an $\eta_+$ pair. This decay may
be detectable when measuring Higgs BRs at a future linear collider.

The quark Yukawa Lagrangian with $\eta$'s present is
\begin{align}
\LL_Y = & \;
  \lambda_1 \overline\chi_{1,R} \chi_{1,L} F \xi_u^3
+ \lambda_2 \overline\chi_{2,R} \chi_{1,L} F \xi_u^3
+ \lambda_3 \overline\chi_{3,R} \chi_{1,L} F \xi_d^3
\notag \\ &
+ \lambda_1 \sqrt{2} \ii (\overline\chi_{1,R} h_u^\dagger q_L) \xi_u^3
- \lambda_2 \sqrt{2} \ii (\overline\chi_{2,R} h_u^\dagger q_L) \xi_u^3
- \lambda_2 \sqrt{2} \ii (\overline\chi_{3,R} h_d^\dagger q_L) \xi_d^3 
\; .
\end{align}
Ignoring the coupling $\lambda_1$ (light quark limit) as well as
$\lambda_d$ (for the down-type quarks)~\cite{Csaki2}, and inserting
the vacuum expectation values of the Higgs fields and keeping at most
trilinear terms, we obtain the fermion mass matrix
\bq
(\overline{\chi}_{1,R}, \overline{\chi}_{2,R}, \overline{\chi}_{3,R})
\begin{pmatrix}
  0 & 0 & 0  \\
  \lambda_2 v \cos\beta \cos\gamma \xi_u^3 & \lambda_2 F \xi_u^3 & 0 \\
  \lambda_3 v \sin\beta \sin\phi \xi_d^3 & 0 & \lambda_3 F \xi_d^3
\end{pmatrix}
\begin{pmatrix}
  t_L \\ \chi_{1,L} \\ \chi_{2,L}
\end{pmatrix} \, .
\eq
Here the additional angles $\gamma$ and $\phi$ parameterize different
scales within different Higgs multiplets.  One linear combination of
$\overline{\chi}_{2,R}$ and $\overline{\chi}_{3,R}$ is the
right-handed top quark, while the orthogonal linear combination mixes
with $\chi_{1,L}$ and $\chi_{2,L}$ to give two heavy quarks with
masses of the order $F$.  Here, the physically relevant rotation is
between the left-handed fermions.  The resulting structure of the
$\eta$ couplings is determined by the general rules given in
Sec.~\ref{sec:littlest}.  Regarding the couplings, the $\eta_+$ is
something like a pseudoscalar Higgs rendered ultralight by a sort-of
Little see-saw mechanism between the scales $v$ and $F$. In contrast,
the $\eta_-$ is a pseudo-axion in the pure sense, with similar
properties as described in the previous section.

The extreme complexity of this model is manifest.  We therefore do not
try to examine its phenomenology in detail.  Rather, we wish to point
out generically that the addition of more heavy quark and lepton
partners can multiply the partial widths of the $\eta$ to $gg$ and
$\gamma\gamma$, which would in general increase the
$gg\to\eta\to\gamma\gamma$ rate at hadron colliders, discussed in the
next section.


\section{Phenomenology}
\label{sec:pheno}

We concluded in the previous section that the $\eta$ coupling
structure to SM fermions and gauge bosons resembles the couplings of a
CP-odd Higgs boson.  Compared to standard two-Higgs doublet model
interactions, however, there are three important differences: (i) all
couplings to SM particles are suppressed by a common factor $v/F$,
which reduces direct production rates by $(v/F)^2$, although for
$gg\to\eta$ at hadron colliders this may be compensated by the
presence of multiple heavy particles running in the loop; (ii) the
$Tt\eta$ coupling is not suppressed and can be similar in magnitude to
$Tth$, altering the phenomenology of the new heavy quarks from that
expected by strict $SU(2)$ symmetry; (iii) the $ZH\eta$ interaction
is not allowed in all models.


\subsection{Production related to heavy {\boldmath $T$} quarks at 
hadron colliders}

The pseudo-axion $\eta$ has essentially Yukawa couplings to the SM
fermions, which are large for the top quark.  Thus, one would
anticipate that associated $t\bar{t}\eta$ production would occur,
analogous to top-pion~\cite{Leibovich:2001ev} or two-Higgs doublet
model pseudoscalar production~\cite{Kominis:1994fa} but the $(v/F)^2$
suppression factor renders this channel useless by trivial comparison
to the SM Higgs.  While the rate for $T\bar{T}$ is {\it not}
suppressed by $(v/F)^2$, it is unfortunately hugely phase-space
suppressed: the $T\bar{T}$ rate is already sub-fb level.

However, the $T$ quark can decay to $t\eta$, often with significant BR
and sometimes dominantly.  At a minimum the $T$ quark phenomenology is
changed in a non-trivial way, since the $\eta$ does not participate in
$SU(2)$.  Because the coupling is highly model-dependent, and not even
completely defined in some models, predictions rapidly become
exercises in having too many free parameters.  On the other hand,
precisely measuring all BRs, both to the ``standard'' final states
$tH$, $tZ$, and $bW$~\cite{PPP03}, and to $t\eta$, could provide deep
insight into the full structure of the model that is realized in
nature, if the Little Higgs mechanism is discovered.

Heavy $T$ quarks can be produced at hadron colliders singly or in
pairs, analogous to SM $t$ quark production.  However, in general
the single-$T$ rate is larger than that for $T\bar{T}$ because of the
enormous phase-space suppression in heavy pair production.
Consequently, only single-$T$ phenomenology has been studied
seriously~\cite{Azuelos:2004dm}.  This study found that even observing
TeV~scale $T$ quarks in a clean channel is difficult with large
luminosity, and the final state $tH$ may be impossible except at the
luminosity-upgraded SLHC~\cite{SLHC}.  While the single-$T$ production
rate is different in the $\mu$ model compared to the Littlest Higgs
model that Ref.~\cite{Azuelos:2004dm} investigated, it is similar.
Because of the extremely complicated final states requiring multiple
large, complicated backgrounds, we do not attempt to study any of
these potential signals in detail here, deferring this to a later
publication~\cite{Kilian:2006eh}.  We do, however, outline some general
features one can expect in the various models.

For the Littlest Higgs case, EW precision data already limit $F\gtrsim
4$~TeV, in which case single-$T$ production rates are fewer than 10
events per 300~fb$^{-1}$.  This is far too few events to utilize,
although at the upgraded SLHC~\cite{SLHC} it might be feasible to
observe the $T$ quark in $Wb$ decays~\cite{Azuelos:2004dm}.  The same
study showed that the decay $tH$ was barely feasible for $m_T=1$~TeV,
so by quick comparison of rates we predict it would be hopeless to see
either the $tH$ or $t\eta$ modes of a Littlest Higgs even at SLHC.
Unfortunately, while the case $\lambda_1\sim\lambda_t$ for Littlest
Higgs makes BR($t\eta$) dominant, for this parameter choice $M_T$ is
simultaneously driven well above the scale $F$: for $F=4$~TeV, this
region of $\lambda_1$ yields $M_T\sim 20-40$~TeV, so that the
$T\bar{T}$ final state is not even accessible at LHC.  Estimates of
the signal cross section alone suggest it would not even be accessible
at a 200~TeV VLHC.  However, it is worth investigating the non-extreme
case of $\lambda_1\sim\lambda_2$ for a VLHC, which is more likely in
any case from the point of view of fine-tuning of the model and the
$T$ quark's role in cancelling the Higgs mass quadratic divergence
coming from the SM $t$ contribution.

The situation is different for $\eta$ in the $\mu$ model, however.
For a type II model, EW precision data do not constrain the scale of
new physics all that much: $F_{1,2}=0.5,2$~TeV is in good agreement
with current data.  For the allowed range $\mu\lesssim 150$~GeV, $m_H$
ranges from 450 down to 120~GeV, $m_\eta$ from 10 up to 310~GeV,
BR($T\to t\eta$) is roughly 1/3, and BR($\eta\to b\bar{b}$) is from
$90\%$ to $27\%$.  The heavy $T$ quark mass remains fixed (here,
990~GeV) so the rate to a triggerable final state is often large.  The
main issues will be that $m_\eta$ is large enough to avoid the large
QCD continuum $b\bar{b}$ contribution, and that it also does not
overlap the Higgs signal.  Due to finite $m_{b\bar{b}}$ resolution,
this means that the resonances will have to be separated by about
50~GeV to be separately visible.

In the $\mu$ model there is also the interesting prospect that $\eta$
can decay to $ZH$, if its mass is large enough, typically
$m_\eta\gtrsim 300$~GeV.  This would allow for the extremely unique
decay signature $T\to t\eta\to tZH$, which is admittedly an extremely
complicated final state with low efficiency for identification, but
also has much smaller backgrounds than any of the standard $T$ quark
final states, or $T\to t\eta\to tb\bar{b}$.  For an $\eta$ mass above
the $t\bar{t}$ threshold $ZH$ remains the dominant decay mode, with a
branching ratio of three quarters, while $t\bar{t}$ is nearly the
remaining quarter, so the $ZH$ signal likely remains feasible for
large $m_\eta$.


\subsection{Direct production at hadron colliders}

Another possible production mechanism, which appears to be more
interesting than that from $T$ decays, is gluon fusion, which proceeds
via the axial $U(1)_\eta$ anomaly, analogous to the $\eta^\prime$ anomaly
of $U(1)_A$ in QCD.  In general, the more heavy fermions in the model,
the more the suppression factor $v/F$ is overcome in the loop-induced
coupling to gluon pairs, and thus the higher the production rate.  As
in SM Higgs phenomenology, the only viable decay channels at the LHC
will be weak boson final states, of which $\gamma\gamma$, similar to
$gg\to H\to\gamma\gamma$~\cite{atlas_tdr,cms_tdr}, offers the best
prospect due to high efficiency from not having subsequent decays.

\begin{figure}\begin{center}
\includegraphics[scale=.9]{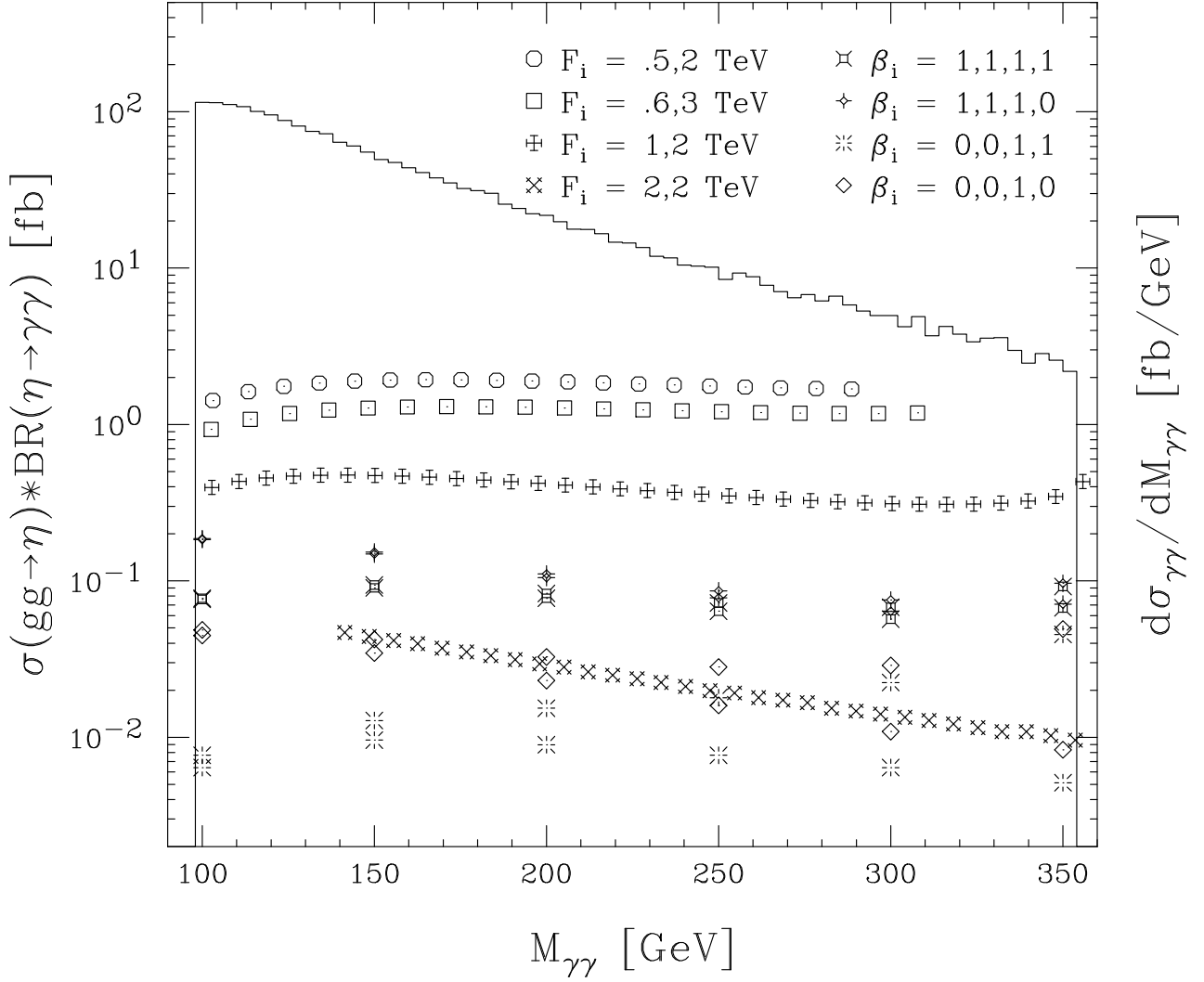}
\caption{Diphoton signal for $gg\to\eta\to\gamma\gamma$ at the LHC, 
shown as individual points which reflect the total signal cross
section for each parameter space choice.  The symbols for the Littlest
Higgs points are shown in the plot legend along with the chosen
$\beta_i$ sets.  The symbols for the $\mu$ model points are shown with
their corresponding choices for $F_1,F_2$.  Two data points are shown
for each value of $m_\eta$ in the Littlest Higgs model, representing
the extremal allowed values of $\lambda_1$: $\lambda_t$ and
$\sqrt{2}\lambda_t$.  We have applied the known NLO K-factors to the
signal cross sections.  The continuum diphoton invariant mass
distribution is shown as a differential cross section histogram with
4~GeV bins.  It includes the direct photon, 1-fragmentation and
2-fragmentation contributions at NLO.  All samples include kinematic
cuts as described in the text, and an ID efficiency factor
$\epsilon_\gamma=0.8$ for each photon.}
\label{fig:gg2gamgam}
\end{center}\end{figure}

We calculate the rates of $gg\to\eta$ production in the Littlest Higgs
and $\mu$ models for a range of allowed parameter choices.  Our
results are shown in Fig.~\ref{fig:gg2gamgam}, where we plot both the
NLO signal absolute cross sections and differentially bin the
continuum diphoton background, which includes the direct, 1- and
2-fragmentation contributions at NLO~\cite{diphox}~\footnote{Diphox
NLO continuum background distributions courtesy of Thomas Binoth.}.
We have applied kinematic cuts of $p_T(\gamma)>40$~GeV and
$|\eta_\gamma|<2.5$, as well as an efficiency factor of
$\epsilon_\gamma=0.8$ for each photon to be identified in the
detector.

For the Littlest Higgs model, we select the minimum allowed scale
choice, $F=4$~TeV, four choices of $\beta_i$ which reflect various
$gg\eta$ and $b\bar{b}\eta$ coupling strengths, and we consider the
mass range $100<m_\eta<350$~GeV.  Unfortunately, these signals appear
to be invisible everywhere.  The most optimistic point is for the case
$\beta_i=[1,1,1,1]$ just below the $t\bar{t}$ threshold, where at the
SLHC with 3000~fb$^{-1}$ the signal might be about $4\sigma$.  Above
350~GeV, the $\eta$ would decay for all practical purposes only to
$t\bar{t}$ final states, which would be lost in the $t\bar{t}$
continuum that is more than two orders of magnitude larger.  The other
choices of $\beta_i$ shown exhibit qulitatively different, but rather
moot, behavior.  For $\beta_{1,2}=0$, the cross sections are typically
an order of magnitude smaller than for the choice $\beta_{1,2}=1$, as
$\beta_1$ directly affects the $t\bar{t}\eta$ coupling.  As $m_\eta$
increases, the ratio $\lambda_1/\lambda_2$ becomes more important.  As
expected, the cross sections are larger for $\beta_3=0$ compared to
non-zero values, as the partial width to $b\bar{b}$ goes to zero,
allowing for a larger BR($\gamma\gamma$).  Negative values for
$\beta_{0,1}$, which result in dominant $T\to t\eta$ decays as
discussed previously, do not result in an enhanced signal here.  The
reason is that the $t\bar{t}\eta$ coupling remains approximately the
same size, but turns negative, resulting in a cancellation between the
$t$ and $T$ loops.  The choice $\beta_i=-2,-2,0,1$ produces signal
cross sections approximately the same size as the case
$\beta_i=1,1,1,0$, so obviously one can search for optimistic cases.
However, since the Littlest model does not define the $\beta_i$, nor
even allow for educated guesses, we cannot speculate further.

The situation in the $\mu$ model is somewhat more positive.  Again in
Fig.~\ref{fig:gg2gamgam} we show the signal as absolute cross
sections, for several choices of $F_1$ and $F_2$.  The $gg\eta$
coupling is comparatively much larger than in the Littlest model,
first because there is no $1/\sqrt{5}$ hypercharge embedding factor
penalty, and second there is the enhancement from the additional heavy
states running in the loop.  Thus, for the Golden Point, the cross
section is more than an order of magnitude larger than the cases shown
for the Littlest Higgs model.  Interestingly, as a function of
$m_\eta$ the product $\sigma_\eta\times$BR($\gamma\gamma$) is nearly
flat.  This is due to the coincidental cancelling of dropoff in
production cross section as $m_\eta$ grows, with the rise in partial
width of $\eta\to\gamma\gamma$.  As $F_1$ is taken to be closer to
$F_2$, the cross section becomes considerably smaller, and for
$F_1=F_2$ it is more than two orders of magnitude smaller.  This is
because the $t\bar{t}\eta$ coupling goes to zero as $F_1\to F_2$,
which not only reduces the production cross section but drives
BR($\gamma\gamma$) to extremely small values.

For the Golden Point at LHC, 300~fb$^{-1}$ would give a $7\sigma$
signal for the highest-mass point allowed, and the signal drops below
$5\sigma$ for about $m_\eta=240$~GeV.  The SLHC reach would extend
down to about $m_\eta=130$~GeV.  For the case $F_{1,2}=0.6,3$~TeV, the
signal is barely $5\sigma$ for $m_\eta=300$~GeV, while at the SLHC one
could discover the $\eta$ in this channel down to about
$m_\eta=160$~GeV.  At the upper end of the $m_\eta$ ranges, the
signal-to-background ratio is a decent $S:B\sim 1/10$, while at the
lower end of the accessibility range it drops to $S:B\sim 1/50$, about
what the SM $H\to\gamma\gamma$ inclusive search would experience.  If
$F_{1,2}=1,2$~TeV, the signal degrades further, with no observability
at LHC.  The SLHC, however, could have limited access, for about
$m_\eta\gtrsim 320$~GeV.  The $F_1=F_2$ cases would always be
invisible.

While we do not work out the Simple Group model case, we can expect that
the signal would in general be a factor of 2 larger than in the
$\mu$ model, based on counting the number of heavy quarks that run in
the loop.  However, the coefficients of the
$t\bar{t}\eta$,$Q\bar{Q}\eta$ couplings may have non-trivial
coefficients, so this is not a rigorous prediction.

A final note is that while we have included the NLO 1- and
2-fragmentation contributions to the diphoton continuum background, we
have not included other sources of fake photons from jets, which
typically are a $20-40\%$ effect, depending on ultimately how well
ATLAS and CMS can reject these events.  We have also not included the
NNLO signal contributions, which are another $20\%$ or so.  Since we
further apply only K-factors for the signal with simple kinematic
cuts, details such as $\gamma\gamma$ recoil are left out.  Yet our
results should still be regarded as conservative and optimistic since
we use 4~GeV bins, which are slightly larger than necessary for
diphoton final states.  Our results are therefore a reasonable
estimate of the reach of the LHC in these Little Higgs models with
pseudo-axions at the weak scale.


\subsection{Pseudo-axion detection at a Linear Collider}

\begin{figure}[t!]\begin{center}
\includegraphics[scale=.85]{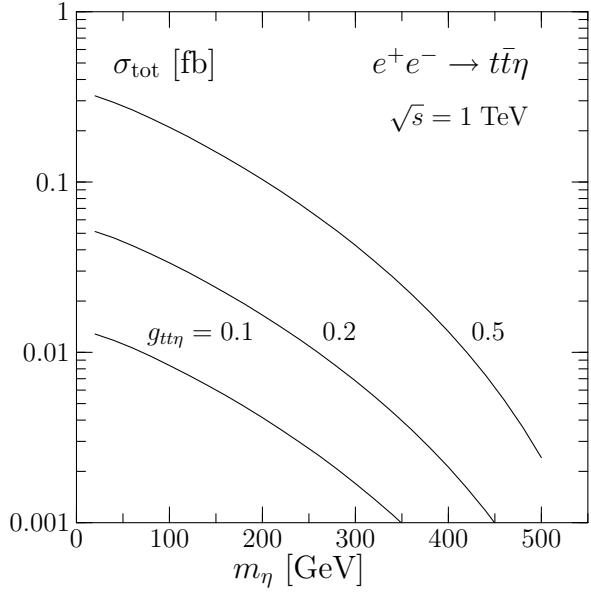}
\vspace{-2mm}
\caption{The $t\bar{t}\eta$ cross section at a 1~TeV linear collider, 
for three different values of $g_{t\bar{t}\eta}$.}
\label{fig:tte}
\vspace{-3mm}
\end{center}\end{figure}
\begin{figure}[t!]\begin{center}
\includegraphics{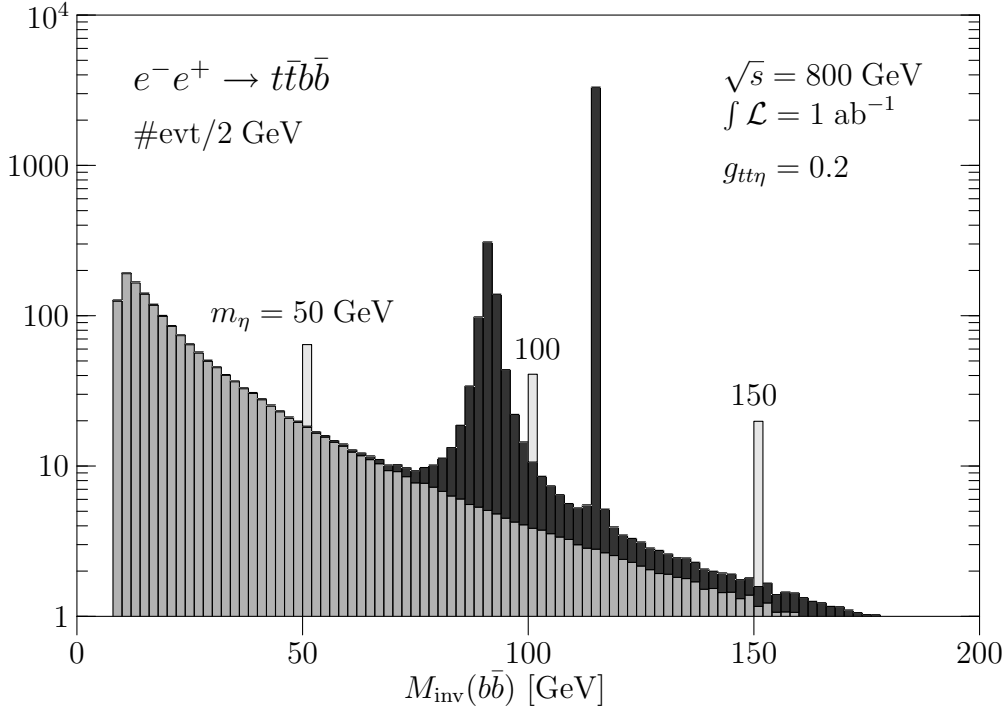}
\caption{Invariant $b\bar{b}$ mass distribution for 
$e^+e^-\to\bar{t}t\bar{b}b$, showing the expected number of events in
$2\;\GeV$ bins for $1\;\ab^{-1}$ of data.  Medium gray is the QCD
contribution, dark gray is EW with $m_H=115\;\GeV$.  The light-gray
spikes are $\eta$ signals for three values of $m_\eta$ with
$g_{tt\eta}=0.2$.  We do not include finite detector resolution
effects, so e.g. the $m_\eta=150$~GeV signal is marginal.}
\label{fig:tte-bg}
\vspace{-3mm}
\end{center}\end{figure}

At an $e^+e^-$ collider, some models will allow for the possibility of
$Z^*\to H\eta$, but we do leave this investigation for a later
publication~\cite{Kilian:2006eh}.  Here we estimate the possibility of
producing $\eta$ in $t\bar{t}\eta$ associated production.
Unfortunately, even for the analogous $t\bar{t}H$ channel, the cross
section is at most a few fb.  For $\eta$, the coupling to top quarks
$g_{tt\eta}$ is proportional to $v/F$, so the rate is
$(v/F)^2$-suppressed.  On the other hand, the rate is enhanced if
$m_\eta$ is significantly below $m_H$.  Fig.~\ref{fig:tte} shows the
total cross section for this process for three choices of
$g_{t\bar{t}\eta}$.

In the mass range below $2m_t$ the $\eta$ decays with significant BR
into a $b\bar{b}$ pair.  The most important background is
$t\bar{t}b\bar{b}$ production.  In Fig.~\ref{fig:tte-bg} we show the
invariant $b\bar{b}$ mass distribution for signal and background, for
three choices of $m_\eta$ and fixed coupling to top quarks.  This
plot, which was calculated using the programs of
Refs.~\cite{Whizard,madgraph} shows the difficulties in detecting this
final state: for low $\eta$ masses, the signal cross section is
sizable but the QCD background is rather large.  If the $\eta$ mass is
close to the $Z$ or Higgs masses, the EW background is oberwhelming.
For larger $m_\eta$ values which are well separated from the $Z$ and
Higgs, the production cross section rapidly decreases.  In any case,
the experimental analysis of the $t\bar{t}b\bar{b}$ final state is
nontrivial~\cite{ttH}, and the achievable resolution in the invariant
$b\bar{b}$ mass (with correct identification of the $t\bar{t}$ pair)
determines the sensitivity for detecting a pseudo-axion in this final
state.

In principle, the process $e^+e^-\to\gamma^*\to\eta\gamma$ is possible
with the help of the anomaly coupling, but it turns out that the
$b\bar{b}\gamma$ background is generally too large for this to be
viable.


\subsection{Pseudo-axions at a photon collider}

\begin{figure}[t!]\begin{center}
\includegraphics[scale=.9]{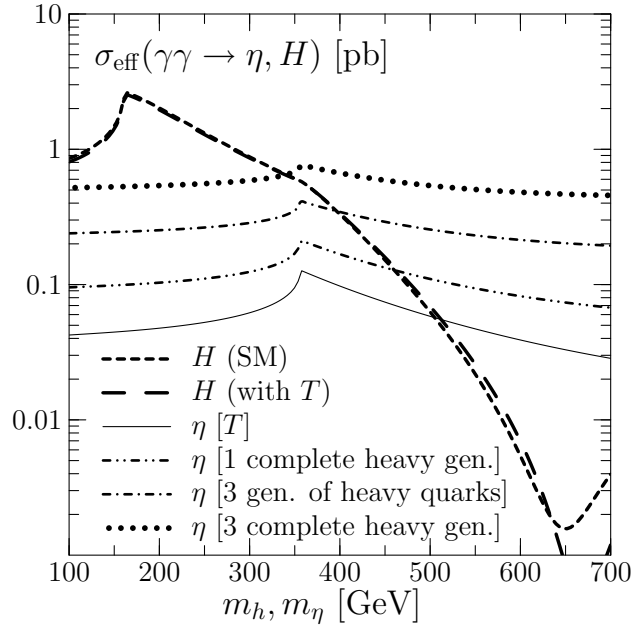}
\caption{Total cross section for resonant $\eta$ production at a future 
photon collider, with different anomaly coefficients as indicated by
the number of non-SM particles with masses $\sim 1\:\TeV$ in the
loop. For comparison, Higgs production cross sections are shown for
the SM as well as SM plus a heavy $T$ quark.}
\label{fig:eta-prod}
\vspace{-5mm}
\end{center}\end{figure}

Especially for Higgs precision measurements, a high energy photon
collider is expected to be operated at a future linear collider, by
Compton backscattering laser photons off the electron beam.  At such a
machine the Higgs boson(s) as well as pseudoscalar Higgs bosons can be
produced as $s$-channel resonances~\cite{phocoll.A}.  Pseudo-axions
could be produced the same way.  The effective cross sections for
resonant pseudo-axion production with different choices of the anomaly
factors (i.e., different numbers of particles in the loop and/or
different coupling factors) are shown in Fig.~\ref{fig:eta-prod}.
Since all strongly-coupled particles running in the loop are much
heavier than the pseudo-axion itself, the cross section remains
constant over a wide range of $m_\eta$, while the Higgs boson shows
the well-known maximum at the $W$ threshold as well as the destructive
interference between gauge boson and fermion loops for masses around
$650\,\GeV$.

\begin{figure}[t!]\begin{center}
\includegraphics[scale=.85]{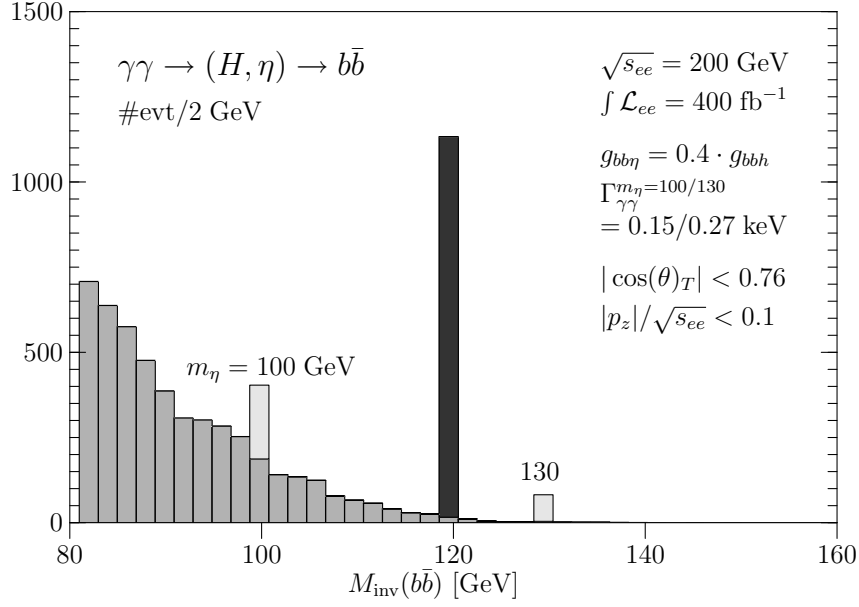}
\caption{Invariant $b\bar{b}$ mass distribution for 
$\gamma\gamma\to b\bar{b}$, showing the expected number of events in
$2\;\GeV$ bins for $400\;\fb^{-1}$ of data.  The light gray spikes are
$\eta$ signals for various $m_\eta$ with $g_{bb\eta}=0.4 g_{bbh}$.  We
also show the SM Higgs signal for $m_H=120\;\GeV$ for comparison (in
the $\mu$ model, $m_H\sim 420\ldots 440\,\GeV$).  Applied kinematic
cuts are shown in the figure.  We do not include finite detector
resolution effects.}
\label{fig:aabb}
\vspace{-3mm}
\end{center}\end{figure}
\begin{figure}[t!]\begin{center}
\includegraphics{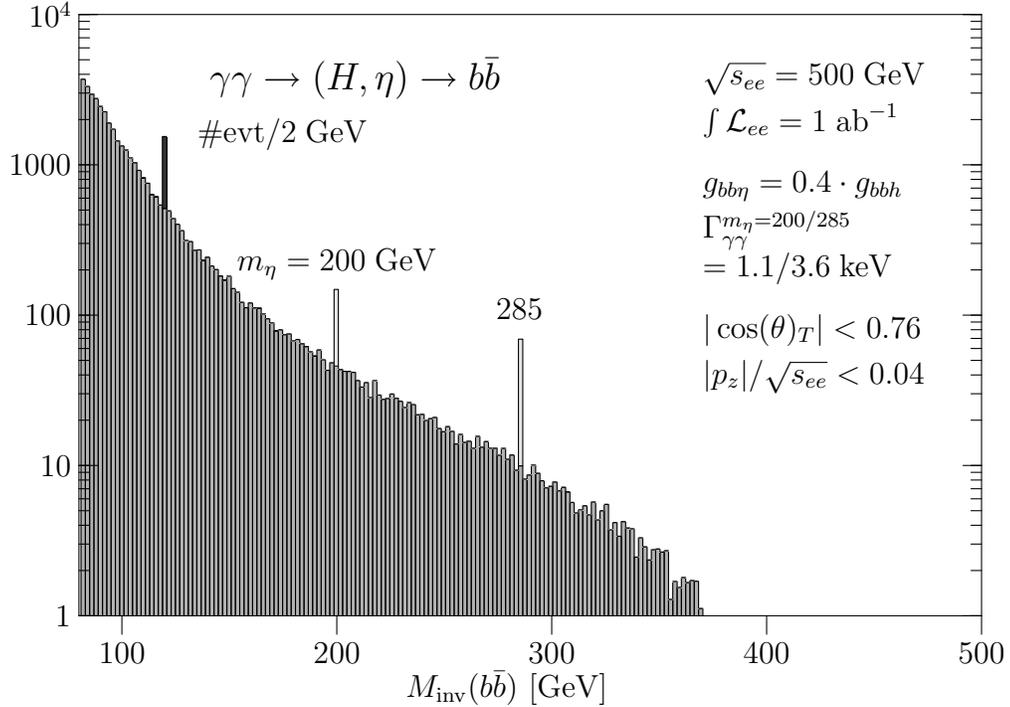}
\caption{Signals for heavy pseudo-axions at the photon collider for 
$m_\eta=200,285\,\GeV$. The values for coupling constants and cuts are
given in the figure. For comparison, the signal for a $120\,\GeV$
Higgs is shown (in the $\mu$ model the corresponding Higgs masses in
the two cases are $368/230\,\GeV$, respectively.)}
\label{fig:heavygg}
\vspace{-8mm}
\end{center}\end{figure}

The decay mode $\eta\to b\bar{b}$ is typically large (although not
necessarily dominant), and the $\eta$ manifests itself as a sharp
spike in the $b\bar{b}$ invariant mass spectrum.  Fig.~\ref{fig:aabb}
shows the $b\bar{b}$ invariant mass spectrum at a 200~GeV photon
collider, which is optimal for studies of a light Higgs boson.
Experimentally, the discovery is challenging since the anomaly factor
must not be too small and the scale $F$ should not be too high in
order for the $\eta$ signal to be clearly visible.  For larger $\eta$
masses the situation is almost the same as for the linear collider:
the background falls rapidly with increasing $m_{b\bar{b}}$, but the
signal cross section falls at a similar rate.  Fig.~\ref{fig:heavygg}
shows the peaks in the $m_{b\bar{b}}$ spectrum for larger $\eta$
masses.  For very heavy pseudo-axions, $m_\eta\sim 350\,\GeV$, the
$t\bar{t}$ channel opens up and one can look for peaks in this
spectrum.  Figs.~\ref{fig:aabb} and \ref{fig:heavygg} were produced
with the programs of Refs.~\cite{Omega,Circe}.

M\"uhlleitner {\em et al.}~\cite{phocoll.A} studied the case of a
pseudoscalar Higgs boson in a certain region of MSSM parameter space,
in which the Higgs pseudoscalar $A$ has exactly the same total width
and BR($b\bar{b}$) as the $\mu$ model $\eta$ at the Golden Point. They
considered finite $b\bar{b}$ resolution and other effects, such as
detector smearing of the signal, and found a significant pseudoscalar
signal at the photon collider.


\section{Conclusions}

If a Little Higgs scenario is realized in nature, Higgs bosons at the
EW scale are typically accompanied by new gauge-singlet pseudoscalar
particles $\eta$ that are associated with the extra spontaneously
broken $U(1)$ symmetry groups.  If these abelian subgroups are gauged,
the particles are absorbed as the longitudinal components of extra
$Z'$ bosons.  At future colliders, such (heavy) vector resonances can
be detected by standard methods.  Their indirect effects on the
existing EW precision data already constrain the parameter space of
Little Higgs models.

Therefore, we have considered the alternative case that at least one
$U(1)$ group is ungauged, so that the associated NGB is physical.  To
avoid the limits on light axions, we require the presence of explicit
$U(1)$ symmetry breaking terms which give the would-be axion a mass in
the EW range.  In particular models, some of the pseudo-axion masses
are calculable, while in other models (such as the Littlest Higgs
model) $m_\eta$ is undetermined.

Detecting these particles would be an important test of the Little
Higgs model structure.  At the LHC, one can perhaps search for them in
decays of the heavy top quark partners $T\to t\eta$ and subsequent
decay $\eta\to b\bar{b}$, or possibly $\eta\to gg$ for extreme
parameter choices or at an upgraded LHC or a future, higher-energy
hadron collider.  In some models, the decay $\eta\to ZH$ is open,
giving rise to extremely distinctive but probably small signals.  In
either case, $T\to t\eta$ decay channels are extremely complicated and
their exact utility will have to await further detailed work.  Our
work has been to point out that the presence of pseudo-axions
generally alters the phenomenology of $T$ quarks, often significantly
and in some cases to the point of domination.

Our much more interesting result is the prospect of searching for
pseudo-axions in direct production at hadron colliders, $gg\to\eta$,
with subsequent decay to photon pairs, in the spirit of Higgs boson
searches.  While observation does not appear to generally be possible
in the Littlest model without fine-tuned choices of the fermion
$U(1)_\eta$ charges, in the $\mu$ model the $\eta$ would be visible at
the LHC for larger values of $m_\eta$, and over significant parameter
space at the luminosity-upgraded SLHC.  The general feature of direct
production is that the more heavy quark partners of the model, the
stronger the production rate; e.g. one would expect even larger rates
in the Simple Group model.  Unlike pseudoscalars in supersymmetry, the
BR($\gamma\gamma$) is not suppressed by fermion couplings enhanced by
$\tan\beta^2$ dominating the total decay width ($\tan\beta$ is
typically restricted to large values in supersymmetry scenarios).
Instead, in general the $b\bar{b}\eta$ couplings are suppressed
relative to Higgs-like couplings, resulting in an often sub-dominant
BR relative to the $gg$ final state, and a non-trivial
BR($\gamma\gamma$) even up to the $ZH$ or $t\bar{t}$ thresholds.

The production channels at a Linear Collider ($e^+e^-\to
t\bar{t}\eta$) or at a photon collider ($\gamma\gamma\to\eta$) are
also promising, and we have presented results for a few cases.  While
the backgrounds to these processes appear manageable, the expected
signal rates are low, so high luminosity and a sophisticated
experimental analysis will be necessary to confirm the presence of a
low-mass pseudo-axion.

The pseudoscalar's nature would easily be distinguished from a
standard Higgs boson by the absence of the $WW$ and $ZZ$ fusion
channels, which are well-known to work over an extremely large mass
range~\cite{WBF}, as well as ``standard'' decay channels such as
$gg\to\eta\to W^+W^-,ZZ$~\cite{Dittmar:1996ss,atlas_tdr,cms_tdr}.

To distinguish $\eta$ from a pseudoscalar Higgs in more conventional
two-doublet models or other extended Higgs sectors, we would have to
identify it as a gauge singlet.  We anticipate that observing all
possible $T$ decay modes would aid this, but it probably would require
an $e^+e^-$ collider which can cover the parameter space as well as
prove the absence of charged and additional CP-even neutral partners.


\subsection*{Acknowledgements}
We thank K.~Desch, R.~Harlander, G.~Hiller, T.~Ohl, M.~Peskin, and
A.~Pierce for useful discussions, Martin Schmaltz and Tim Tait for
critical reviews of the manuscript, and Thomas Binoth for providing us
with Diphox NLO diphoton continuum predictions for the LHC.  W.K. is
grateful for the hospitality of the SLAC Theory Group, and D.R. thanks
the KITP for its hospitality, where part of this work was completed.
This research was supported in part by the National Science Foundation
under Grant No. PHY99-07949, the U.S. Department of Energy under grant
No. DE-FG02-91ER40685, and by the Helmholtz-Gemeinschaft under Grant
No. VH-NG-005.  J.R. was also supported by the DFG
Sonderforschungsbereich (SFB) ``Trans\-regio 9 -- Computergest\"utzte
Theoretische Teilchenphysik'' and the Graduier\-tenkolleg (GK)
``Hoch\-energiephysik und Teilchenastrophysik''.


\appendix


\section{Heavy $T$ quark decay partial widths}
\label{sec:decay}

In the Littlest Higgs model, the $T$ partial decay widths $T\to tH$
and $T\to t\eta$ are given by
\begin{align}
\Gamma & =
  \frac{M_T}{32\pi} f(x_t, x_h)  
  [(1 + x_t^2 - x_h^2)(\kappa_L^2 + \kappa_R^2) + 4 \kappa_L\kappa_R x_t]
\end{align}
with the usual definitions $x_i = m_i/M_T$ and the function
\begin{align}
f(x_i, x_j) & = \sqrt{(1-(x_i+x_j)^2)(1-(x_i-x_j)^2)},
\end{align}
where for $H$ and $\eta$ the left- and right-handed couplings are
\begin{align}
  \kappa_L^H    & = \Op\left( \frac{v}{F} \right)
& \kappa_R^H    & = -\frac{s}{c}\frac{m}{v} \\
  \kappa_L^\eta & = i \frac{m}{v}\beta' 
& \kappa_R^\eta & = i \Op\left( \frac{v}{F} \right)
\end{align}
In the limit $v/F\to 0$ where the masses of $H,Z,W$, and $\eta$ can be
neglected compared to $M_T$, $SU(2)$ symmetry relates the partial
decay widths into (longitudinally polarized) vector bosons to the
partial decay widths into a Higgs boson, and the BRs simplify to
\begin{align}
\Gamma(T\to t H) & =
\frac{M_T}{32 \pi}\frac{m^2}{v^2}\frac{s^2}{c^2}
  = \Gamma(T\to t Z^0) = \frac12\Gamma (T\to b W^+) \\
\Gamma(T\to t\eta) & =
\frac{M_T}{32\pi}\frac{m^2}{v^2}\beta^{\prime 2} \; .
\end{align}

For the $\mu$ model, the partial widths of the heavy $T$ quark
(neglecting the $b$ quark mass) are:
\begin{align}
\Gamma(T\to th) &=\; 
  \frac{M_T}{32\pi} f(x_t,x_h) 
  \left[ (1 + x_t^2 - x_h^2) (\kappa_L^{h\,2} + \kappa_R^{h\,2}) 
         + 4 \kappa_L^{h}\kappa_R^{h} x_t
  \right] \\
\Gamma(T\to t\eta) &=\; 
  \frac{M_T}{32 \pi} f(x_t,x_\eta) 
  \left[ (1 + x_t^2 - x_\eta^2) (\kappa_L^{\eta\,2} 
         + \kappa_R^{\eta\,2}) + 4 \kappa_L^{\eta}\kappa_R^{\eta} x_t 
  \right] \\
\Gamma(T\to Zt) &=\; 
  \frac{\sqrt{\lambda(M_T^2,m_Z,m_t)}}{16\pi M_T^3} 
  \biggl\{ (c_V^2 + c_A^2) \Bigl( M_T^2 - m_Z^2 + m_t^2 + \\ & \qquad 
           \frac{(M_T^2 + m_Z^2 - m_t^2)(M_T^2 - m_Z^2 - m_t^2)}{m_Z^2} \Bigr)
            - 6 M_T m_t (c_V^2 - c_A^2) 
  \biggr\} \\
\Gamma(T\to Wb) &=\; 
\frac{\sqrt{\lambda(M_T^2,m_W^2,0)}}{32 \pi  M_T^3} 
      g_{TWb}^2 (M_T^2 - m_W^2) \left(2 + \frac{M_T^2}{m_W^2}\right)
\end{align}
with 
\bq
\lambda(x,y,z) = (x - (y + z)^2)(x - (y - z)^2) \; .
\eq
Yukawa-type couplings are defined as
\bq
\bar\Psi \gamma^\mu (c_V - \gamma^5 c_A) \Psi, \qquad
g \bar\Psi \gamma^\mu \frac12 (1 - \gamma^5) \Psi \; .
\eq
%


\section{Loop-induced couplings}
\label{sec:loop}

Little Higgs pseudo-axions couple to vector bosons at one-loop order
via the usual triangle graphs in Higgs phenomenology.  All fermions
which get their mass by $U(1)_\eta$ breaking run in the loop.  As long
as these fermions are heavy compared to the axion, the loop value is
independent of the heavy-fermion mass.  It depends only on the
triangle anomaly coefficient, i.e., the magnitudes of the effective
$\eta\gamma\gamma$, $\eta W^+W^-$, $\eta ZZ$ and $\eta gg$ vertices,
given by the mixed anomalies of the $U(1)_\eta$ symmetry with the
electromagnetic, EW, and QCD gauge groups, respectively.  We write the
anomaly coefficients $C_i$ as parameters:
\bq\label{anomaly}
\begin{split}
\LL_{\text{anom.}} & =
  \frac{1}{F}\frac{\alpha_s}{8\pi^2}C_g\cdot\eta G_{\mu\nu}\tilde G^{\mu\nu}
+ \frac{1}{F}\frac{\alpha}{8\pi^2}C_\gamma\cdot\eta A_{\mu\nu}\tilde{A}^{\mu\nu}
\\ &\quad
+ \frac{1}{F}\frac{\alpha}{4\pi^2}C_{Z\gamma}\cdot\eta Z_{\mu\nu}\tilde{A}^{\mu\nu} 
+ \frac{1}{F}\frac{\alpha}{8\pi^2}C_Z\cdot\eta Z_{\mu\nu}\tilde{Z}^{\mu\nu}
+ \frac{1}{F}\frac{\alpha}{4\pi^2}C_W\cdot\eta W^+_{\mu\nu}\tilde{W}^{-\,\mu\nu}
\end{split}
\eq
The dual field strengths are normalized by $\tilde{F}^{\mu\nu} =
\frac12\epsilon^{\mu\nu\rho\sigma}F_{\rho\sigma}$ The values of the
anomaly coefficients are model-dependent and given by
\begin{subequations}
\begin{align}
C_\gamma & = \; \sum_f N^f_c Q_f^2
  |F^\eta_{1/2}(4 m_f^2/m_\eta^2)|\frac{g_{\eta f\bar{f}}}{\lambda_f}
\, , \\
C_g      & = \; \sum_q \frac{1}{2} \,
  |F^\eta_{1/2}(4 m_q^2/m_\eta^2)|\frac{g_{\eta q\bar{q}}}{\lambda_q}
\, ,
\end{align}
\end{subequations}
where the mass of the fermion in the loop is parameterized by $m_f =
\lambda_f F$ (for heavy particles $\lambda_f$ and $g_{\eta f\bar{f}}$
are both order one, while for SM particles like the top both are order
$v/F$) and $F_{1/2}^\eta$ is the function defined in~\cite{hunter}. We
are not interested in the operators containing the $W$ or the $Z$ here
because the $\eta$ already has small production cross sections, and
further small BRs to observable (lepton) final states makes these
decays moot.  The relevant couplings are given in Eq.~(\ref{yuk-exp2})
for the Littlest Higgs model and in Eq.~(\ref{muyuk}) for the $\mu$
model.  Unlike for a Higgs scalar, the $W$ loop is unimportant, since
the $\eta$ does not couple to weak bosons at tree level.  Since the
$W$ loop would otherwise contribute with opposite sign to fermion
loops, there is no destructive interference as in the SM.

In contrast, in Little Higgs models the masses of all new heavy
fermions are expected to be non-invariant under the $U(1)_\eta$
symmetries (to make the model natural), hence they all contribute to
the axion-gauge boson interactions with full strength.  In particular,
the Simple Group models discussed above predict heavy partners for all
SM fermions.  Leaving aside the detailed coupling structure, for the
$\eta gg$ coupling the heavy quark triangle diagram value is
approximately multiplied by $N=3$ (the number of quarks) in the $\mu$
model, or $N=6$ in the Simple Group model.  For the coupling to EW
gauge bosons, the heavy partners of leptons and neutrinos have also to
be included in the loop, which we ignore here.  Note also that in
extended scalar sectors in models like the $[SU(4)/SU(3)]^4$ simple
group, there can be enhancement effects by the tangent of a mixing
angle analogous to that of the MSSM.  The loop integrals for the
triangle graphs are found in~\cite{hunter}.

In the Littlest Higgs model, the situation is less certain since we do
not know how many heavy fermions actually exist.  Furthermore, we need
the absolute $U(1)_\eta$ charges of the fermions, while the previously
introduced $\beta$ coefficients are merely differences of $U(1)_\eta$
charges.  Here, we cannot predict the anomaly coefficients but have to
leave them as free parameters.  It is even possible for them to cancel
for randomly chosen integer values of $\beta_0,\beta_1,\beta_2$.  This
is largely irrelevant, however, as we note that the normalization
factor $1/\sqrt{5}$ squared highly suppresses the anomalous partial
widths.  This is peculiar to the Littlest Higgs model and is due to
the hypercharge embedding.  However, judicious choice of the $\beta$
coefficients can compensate for this.

Nevertheless, it is not unreasonable to expect that the factor $v/F$,
which suppresses the $\eta$ couplings compared to the corresponding
CP-even or CP-odd Higgs couplings in the SM or MSSM, is compensated by
the large weight of the heavy-fermion sector in any Little Higgs
model.


\end{document}